\documentclass{aa}  

\usepackage{graphicx}
\usepackage{txfonts}
\def\iso#1{$^{#1}$}
\def\msun{M$_\odot$}

\begin{document}

   \title{The $s$ process in AGB stars as constrained by a large sample of Barium stars}

   \author{B. Cseh\inst{1},         
          M. Lugaro\inst{1,2},
          V. D'Orazi\inst{3},
          D. B. de Castro\inst{4},
          C. B. Pereira\inst{5},
          A. I. Karakas\inst{2},\\
          L. Moln\'ar\inst{1,6},
          E. Plachy\inst{1,6},
          R. Szab\'o\inst{1,6},
          M. Pignatari\inst{7,1,8,9},
          S. Cristallo\inst{10,11}\\
          }

   \institute{
   	$^1$ Konkoly Observatory, Research Centre for Astronomy and Earth 
Sciences, Hungarian Academy of Sciences, \\ H-1121 Budapest, Konkoly Thege M. \'ut 15-17, Hungary
    \\    \email{cseh.borbala@csfk.mta.hu}
    \\$^2$ Monash Centre for Astrophysics, School of Physics and Astronomy, Monash University, VIC 3800, Australia
    \\$^3$ INAF Osservatorio Astronomico di Padova, Vicolo dell’Osservatorio 5, 35122 Padova, Italy
	\\$^4$ Department for Astrophysics, Nicolaus Copernicus Astronomical Centre of the Polish Academy of Sciences, 00-716 Warsaw, Poland
	\\$^5$ Observatorio Nacional, Rua General Jos\'e Cristino, 77 Sao Cristovao, Rio de Janeiro, Brazil
	\\$^6$ MTA CSFK Lendület Near-Field Cosmology Research Group
	\\$^7$ E. A. Milne Centre for Astrophysics, Department of Physics \& Mathematics, University of Hull, HU6 7RX, United Kingdom
	\\$^8$ The NuGrid Collaboration, http://www.nugridstars.org
	\\$^9$ Joint Institute for Nuclear Astrophysics - Center for the Evolution of the Elements, USA
    \\$^{10}$ INAF, Osservatorio Astronomico d'Abruzzo, I-64100 Teramo, Italy 
    \\$^{11}$ INFN-Sezione di Perugia, I-06123 Perugia, Italy
             }

   \date{Received ; accepted }

 
  \abstract
   {Barium (Ba) stars are dwarf and giant stars enriched in elements heavier than iron produced by the $slow$ neutron-capture process ($s$ process). They belong to binary systems where the primary star evolved through the asymptotic giant branch (AGB) phase, during which it produced the $s$-process elements and transferred them onto the secondary, now observed as a Ba star.}
   {We compare the largest homogeneous set of Ba giant star observations of the $s$-process elements Y, Zr, La, Ce, and Nd with AGB nucleosynthesis models to reach a better understanding of the $s$ process in AGB stars.}
   {By considering the light-$s$ (ls: Y and Zr) heavy-$s$ (hs: La, Ce, and Nd) and elements individually, we computed for the first time quantitative error bars for the different hs-element/ls-element abundance ratios, and for each of the sample stars. We compared these ratios to low-mass AGB nucleosynthesis models. We excluded La from our analysis because the strong La lines in some of the sample stars cause an overestimation and unreliable abundance determination, as compared to the other observed hs-type elements.}
   {All the computed hs-type to ls-type element ratios show a clear trend of increasing with decreasing metallicity with a small spread (less than a factor of 3). This trend is predicted by low-mass AGB models where \iso{13}C is the main neutron source.
   The comparison with rotating AGB models indicates the need for the presence of an angular momentum transport mechanism that should not transport chemical species, but significantly reduce the rotational speed of the core in the advanced stellar evolutionary stages. This is an independent confirmation of asteroseismology observations of the slow down of core rotation in giant stars, and of rotational velocities of white dwarfs lower than predicted by models without an extra angular momentum transport mechanism.
   }
   {}

   \keywords{Stars: abundances -- Nuclear reactions, nucleosynthesis, abundances -- Stars: AGB and post-AGB
   }
   \authorrunning{B. Cseh et al.}           
   \maketitle
   

\section{Introduction}

Barium (Ba) stars are a type of chemically peculiar star (G-K giants or dwarfs) belonging to a binary system where the material formed in the interior of the more evolved primary star during the asymptotic giant branch (AGB) phase was transferred onto the companion star. Consequently the less evolved secondary star is enriched in the elements heavier than Fe (including Ba) that were synthesised in the AGB star companion through the $slow$ neutron capture ($s$) process \citep[e.g.,][]{busso99}. As it was first recognised by \citet{Bafirst}, Ba stars also show strong CH, C$_{2}$ molecular bands, indicating enrichment in C. This is in agreement with the idea of mass transfer from an AGB star, since these stars produce also carbon. Further evidence for the AGB origin of the abundances of Ba stars was provided by the discovery of a low-metallicity Ba star rich in F, another AGB product \citep{Alves-Brito11}.
Radial velocity observations confirm this scenario: the large majority of Ba stars show radial velocity variations originating from orbital motion, meaning that a companion star is present in the system \citep[see, e.g.,][]{griffin80,griffin_herbig81,RV2-McC,jorissen_mayor88,mcc_woodsw90,jorissen98}.
The primary star in these systems should now be a white dwarf (WD), as  confirmed in some cases by UV observations \citep[see, e.g.,][]{B-V80,Schindler82,Dominy83,B-V84,Jorissen96,B-V00,Frankowski06}. 
The features of the mass transfer mechanism that produced these systems, most likely wind accretion, however, are still under debate. A main problem is that wind accretion predicts much larger final orbits for the binary systems than those observed in the Ba star population. Solutions include a white-dwarf kick \citep{izzard10}, although this possibility has been disputed by \citet{Milliman15}, and an eccentricity-enhancing mechanism due to tidally enhanced mass loss \citep{bonacic08}. Hydrodynamical simulations of wind mass transfer have been performed by \citet{liu17}, who found that along with increase of the mass ratio of the system, also the mass-accretion efficiency and the accreted specific angular momentum increase. A potential problem is that accretion increases the angular momentum such that the star quickly reaches critical rotational velocity \citep{matrozis17}.

In AGB stars, C and $s$-process elements are produced in the He-rich intershell located between the H-burning and the He-burning shells. The two burning shells are activated alternately: H burning is present most of the time, and it is recurrently interrupted by brief episodes of He burning (thermal pulses, TPs). These produce enough energy at the bottom of the intershell to drive a convective zone within it. Once this convective region disappears, the extended H-rich convective envelope deepens into the He-rich intershell layers, carrying to the stellar surface the C resulting from partial He burning. These recurrent mixing episodes are collectively known as the third dredge-up (TDU). The $s$-process elements are produced via the capture of neutrons on Fe seeds, with neutrons released both during the H-burning phases, within a so-called \iso{13}C ``pocket'' by the \iso{13}C($\alpha$,n)\iso{16}O reaction at temperatures in the order of 100 MK, and during the TPs by the \iso{22}Ne($\alpha$,n)\iso{25}Mg reaction, if the temperature reaches above $\simeq$300 MK \citep{Gallino98,goriely00,busso01,lugaro03,cristallo09}. The resulting abundances are carried to the stellar surface via the TDU. In AGB stars of relatively low mass ($<$ 4--5 \msun) the temperature barely reaches 300 MK, and the \iso{13}C nuclei are the predominant neutron source. For higher masses the \iso{22}Ne neutron source becomes predominant. To generate enough \iso{13}C nuclei in low-mass AGB stars it needs to be assumed that some partial mixing occurs between the base of the convective envelope and the radiative intershell at the deepest extent of each TDU episode, leaving an abundance profile of protons in the radiative He-rich region. The protons react with the abundant \iso{12}C to produce the \iso{13}C pocket. Within this pocket the neutron exposure $\tau$, i.e., the time-integrated neutron flux, is relatively large, of the order of 10$^{-1}$ to 1 mbarn, resulting in strong enhancements (up to 10$^{3-4}$ times the initial abundances) of the nuclei heavier than Fe. These are mixed inside the TP convective region and then carried to the stellar surface by the following TDU. See \citet{herwig05}, \citet{straniero06}, \citet{kaeppeler11}, and \citet{karakas14} for recent reviews on AGB stars and the $s$ process.

Barium stars can provide us stringent constraints on the $s$ process in AGB stars because we can perform a more straightforward, accurate, and precise derivation of the abundances of different elements on their surfaces than on the AGB stars themselves. The temperatures of Ba stars (from over 4000K up to 6500K) are higher than those of late AGB stars ($\simeq$ 3000--4000K), consequently, the spectra of Ba stars are easier to model than those of AGB stars because there are less molecules. Furthermore, the atmospheres of AGB stars are characterised by strong dynamical processes, such as pulsations and mass loss, as well as dust formation, which makes their modelling challenging \citep{abia02,perez17}. 
Ba stars cover a range of metallicity roughly from solar to 1/10$^{\rm th}$ of solar, which makes them the higher-metallicity equivalent of $s$-process enriched CH and carbon-enhanced metal-poor stars (CEMP-$s$) in the halo \citep[e.g.,][and references therein]{lucatello05,masseron:10,bonifacio:15,cristallo16,yoon:18}. 

Up to a couple of years ago, observations of Ba stars were limited, both in number and in terms of the self-consistency of the spectral analysis. 
\citeauthor{AB2006} (\citeyear{AB2006, AB2006_2}) presented a self-consistent sample of 26 Ba dwarfs and giants observed with the FEROS spectrograph \citep[R$=$48000][]{kaufer99}. \citet{Smiljanic07} compared normal giants and roughly 10 Ba stars with different $s$-process enhancements. The spectra were also taken with the FEROS spectrograph with high S/N ratio ($\approx$500--600). 
The number of known Ba stars grew with the discovery of 12 metal-rich Ba giants based on a high resolution spectroscopic survey of 230 stars \citep{Pereira11}. \citet{Yang2016} derived atmospheric parameters and abundances for 19 Ba giants with somewhat lower resolution (R$\sim$30000) than the other observations. 
Recently, \citet{Karinkuzhi18a} published an analysis of three Ba stars based on FEROS spectra and \citet{karinkuzhi18b} observed further 18 Ba stars with the HERMES spectrograph \citep{hermes}.
All the mentioned papers observed different $s$-process elements.
The situation changed substantially since the study by \citet{deC} (hereafter deC16), where 182 Ba stars candidates were observed and analysed. Here, we present a detailed comparison between this largest available sample of high resolution spectroscopic observations of Ba stars from deC16 and predictions for the abundances of the $s$-process elements produced in AGB models. The final aim is to derive the implications of these new data on our understanding of the $s$-process in AGB stars.


\section{Sample stars and error calculations}
\label{sec:data}

   \begin{figure}
   \centering
   \includegraphics[width=\hsize]{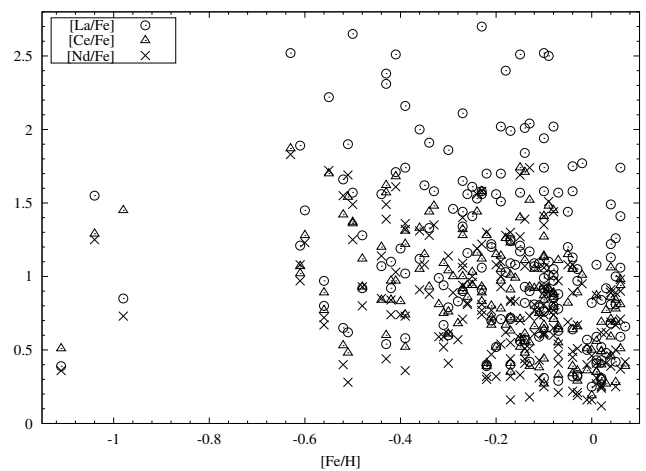}
      \caption{[La/Fe], [Ce/Fe] and [Nd/Fe] values for the 169 sample stars showing the difference between the individual [hs/Fe] abundances.}  
    \label{fig:highla}
     \end{figure}
   
We consider the largest, self-consistent data set of abundances derived from high-resolution observational data of bona fide and possible Ba stars published by deC16. The whole sample of 182 stars is an order of magnitude larger than those available previously (e.g., as mentioned above, \citealt{Antipova2004} with 16 stars, \citealt{AB2006} with 26 stars, and \citealt{Yang2016} with 19 stars) and includes Ba giant stars in a wide range of mass, temperature, and metallicity. The observations were carried out with the FEROS spectrograph \citep[R$=$48000][]{kaufer99}, which covers the spectral region from 3800 to 9200 \AA. The determination of the atmospheric parameters and the abundances of the elements were calculated by measuring the equivalent widths (EW) of the spectral lines. See Sec.~4 of deC16 for more details. 
The $s$-process elements analysed are Y (5 lines), Zr (20 lines), La (5 lines), Ce (10 lines), and Nd (16 lines). The Ba abundances are not available because the Ba lines are very strong in the sample stars (EW typically between 200 and 400 m\AA).  
The selection criteria of deC16 for a Ba star is [s/Fe]\footnote{Throughout the paper we use the standard spectroscopic notation [X/Y]=log(X/Y)$_\star$ $-$ log(X/Y)$_\odot$, where X and Y are abundances by number.} $>$ 0.25, [s/Fe] meaning the average of [Y/Fe], [Zr/Fe], [La/Fe], [Ce/Fe], [Nd/Fe]. Thirteen stars did not match this criteria and we do not include them in our comparison. The final number of Ba stars considered here is thus 169.

To date, extensive use of the so-called ``ls'' and ``hs'' indexes has been made in relation to observations of $s$-process elements. These indexes are calculated as the arithmetic average of the [X/Fe] ratios of the elements belonging to the first (light, ls) and second (heavy, hs) $s$-process peaks \citep[][]{luck:91}, referred as peaks due to their higher abundances in the solar-system $s$-process distribution. Observable elements heavier than Fe that have been considered to belong to the first/second $s$-process peak are Sr, Y, and Zr/Ba, La, Ce, Pr, Nd, and Sm and are characterised by isotopes with or close to neutron magic number of neutrons $N=50$/$N=82$. A third peak is located at Pb with $N=126$. These elements are more produced by the $s$ process relatively to the other elements because their neutron-capture cross sections are significantly lower, by up to orders of magnitude, than those of the other isotopes heavier than Fe. During a neutron exposure these magic neutron isotopes are less likely to capture a neutron; they accumulate, resulting in higher abundances. 

Here, we will not use the $s$-process indexes described above, but consider directly each first and second observed $s$-process peak element separately. The main reason is that the usage of the ls and hs indexes does not allow a straightforward calculation of the uncertainty. When comparing to the models, a typical value of the order of $\pm$0.25 dex has been taken as the error bar for all the stars considered, rather than computed for each individual star. The ls and hs indexes were originally introduced to maximize the information from the spectroscopic analysis and variations in the choice of elements used to calculate them were made to follow the quality of the spectra. However, with the high-resolution spectra and self-consistent sample analysis of deC16 there is a priori no need to maximize information nor to select elements depending on the quality of the spectra. 
Considering elemental ratios directly rather than averaged indexes allows us to significantly improve the error analysis, calculate uncertainties for individual stars, and single out potential issues related to specific elements.
Furthermore, observational studies are able to observe only a number of elements belonging to the first and second peak and the choice of elements used to compute the indexes needed to be adjusted accordingly. Theoretical studies have used different elements to compute ls and hs, see, for example, definitions in \citet{busso01} and \citet{lugaro12}. This is in principle justified because different elements belonging to the same $s$-process peaks are to first order produced in similar amount by the $s$ process relative to their solar abundances. However, it can lead to inconsistencies of the order of 0.1-0.2 dex when comparing results to each other, in particular when considering elements such as Nd, Pr, and Sm whose solar abundances are not predominantly of $s$-process origin \citep{arlandini99,bisterzo11}. 
  
Sixteen out of 169 stars in our sample show unexpectedly high La abundances with [La/Fe]$>$2.0 (Fig.~\ref{fig:highla}). These [La/Fe] values are much higher than the [Ce/Fe] and [Nd/Fe] values. For example, HD 24035 has the highest [La/Fe] with 2.70 dex, while both [Ce/Fe] and [Nd/Fe] are 1.58 dex. Likewise HD 43389 has [La/Fe] $=$ 2.65, [Ce/Fe] $=$ 1.36 and [Nd/Fe] $=$ 1.49. This is not possible to explain by the $s$ process, which by definition predicts that elements belonging to the same $s$-process peak have similar enhancements (see also Sec.~\ref{sec:models}), and we considered the possibility that the La enhancements in these stars may be an observational artefact. We found that they are most likely caused by the very strong 
La lines (EW $\gtrapprox 150$ m\AA) in their spectra. For this reason we decided to exclude La from further analysis. The other hs elements (Ce and Nd) considered by deC16 do not have such strong lines as La and can be used safely as proxies of the second $s$-process peak.
\citet{Smiljanic07} also noted that the uncertainties of the $s$-process elements abundances for some of their sample stars may be underestimated due to the same problem of determining accurate abundances for elements with very strong lines in the spectra of Ba stars.

We calculated individual error bars for each ratio between a second (Ce and Nd) and a first (Y and Zr) peak element by combining the uncertainties coming from the model atmospheres, and from the dispersion of the observed abundances if at least three lines are available for the given element. These calculated errors are likely to be upper limits because most of the stellar parameters are not independent from each other. The steps of our error calculations are the following:

   \begin{enumerate}
   
   \item Calculate the variation as the difference of two elements (X and Y) for $\Delta$T$_{\rm eff}$, $\Delta$log$g$, $\Delta\xi$, $\Delta$[Fe/H], and $\Delta$W$_\lambda$
   from Tables 9, 10, and 11 of deC16. 
   The stars were grouped in these three tables according to three temperature ranges: 5000-5400 K, 4700-4900 K, and 4100-4600 K. The uncertainty was calculated as: 

   $\sqrt{\left( \dfrac{\partial \left[\frac{X}{Y}\right]} {\partial T_{\textrm{eff}}} \right)^2 + \left( \dfrac{\partial \left[\frac{X}{Y}\right]} {\partial \textrm{log}g} \right)^2 + \left( \dfrac{\partial \left[\frac{X}{Y}\right]} {\partial \xi} \right)^2 + \left( \dfrac{\partial \left[\frac{X}{Y}\right]} {\partial \rm[Fe/H]} \right)^2 + \left( \dfrac{\partial \left[\frac{X}{Y}\right]} {\partial \textrm{W}_\lambda} \right)^2}$

    The calculated value for each group was applied to all the stars belonging to the same group. 
   
   \item Calculate the uncertainty coming from the dispersion of the abundances as:
$\sqrt{(\sigma_{\rm obs}/\sqrt{N})_{A}^2+(\sigma_{\rm obs}/\sqrt{N})_{B}^2}$, where A and B are two different elements, $N$ is the number of lines for each element and $\sigma_{\rm obs}$ is the dispersion of the abundances among different lines.

   \item Combine the errors calculated in steps 1 and 2 above by taking the square root of the sum of the squares of the two errors.

   \end{enumerate}

   \begin{figure}
   \centering
   \includegraphics[width=\hsize]{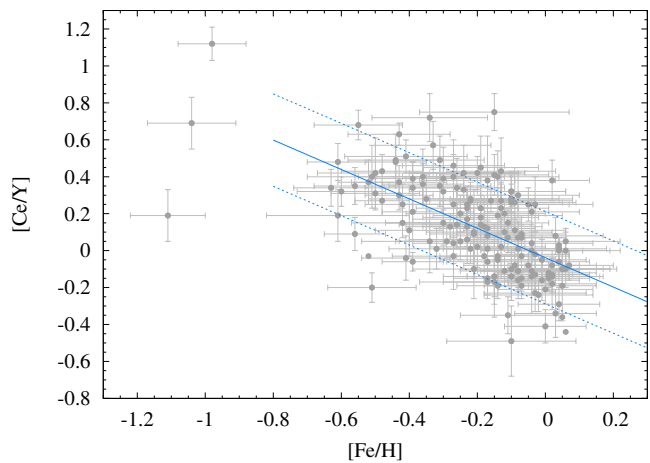}
      \caption{[Ce/Y] values for the 169 sample stars. The dots without error bars represent stars for which there are less than 3 lines for one of the elements. The solid line represents the weighed fit through the data points with [Fe/H] between solar and $-$0.8. The equation of the line is ($-0.038\pm0.021) + (-0.796\pm0.087)\times x$. The dotted lines represent a spread of $\pm$0.25 around the fitted line.
      \label{fig:cey}}
      \end{figure}
 
Fig.~\ref{fig:cey} shows the [Ce/Y] ratios as function of [Fe/H] as an example of the results. The [Fe/H] error for each star is the dispersion given in deC16. The distribution presents: (1) a clear trend of [Ce/Y] increasing by 0.6 dex as [Fe/H]  correspondingly decreases by 0.6 dex (weighed fit solid line); (2) a spread of roughly a factor of three (dashed lines at $\pm$0.25) at any given metallicity, (3) a few ($<$5\%) outliers located with their error bars outside the dashed lines at both higher and lower [Ce/Y] values.      
These results represent a significant improvement with respect to all the previously available data. 
In fact, early studies aimed at understanding the physics of the $s$ process in AGB models using spectroscopic observations \citep{smithlambert88,busso92,busso95,busso01} had to rely on very limited data sets and to consider together all the available observations, not only from Ba stars but also from AGB stars themselves and post-AGB stars \citep[see, e.g., Figure~12 of][]{cristallo11}. However, different samples from different studies are prone to be inconsistent with each other, and further, different types of stars may represent the signature of different physics.


\section{Models}
\label{sec:models}

For comparison to the observational data we consider two sets of models: the FRUITY models available online via the FRUITY database\footnote{http://fruity.oa-teramo.inaf.it/} \citep{cristallo09,cristallo11,cristallo15} and the Monash models \citep{lugaro12,fishlock14,karakas16,karakas18}. These data sets are the most complete available for the $s$-process in AGB stars in terms of stellar mass and metallicity. The FRUITY models are available for 8 metallicities between Z=0.001 and 0.02, covering the whole range of Ba stars. The Monash models are available for 5 metallicities between Z=0.001 to 0.03. 
Models produced by the NuGRID collaboration have also become recently available \citep{pignatari16,battino16,ritter17} for 4 metallicities in the range of interest here: 0.001, 0.006, 0.01, and 0.02. We do not include these models in the following tables but will discuss them and compare with the Monash and FRUITY models where relevant\footnote{Another recent stellar AGB model yield set for different metallicities has been used in a focus study on the Galactic chemical evolution of heavy elements \citep{bisterzo:16}, we do not consider it here since tabulated data are not available.}.

\begin{figure}
   \centering
   \includegraphics[width=\hsize]{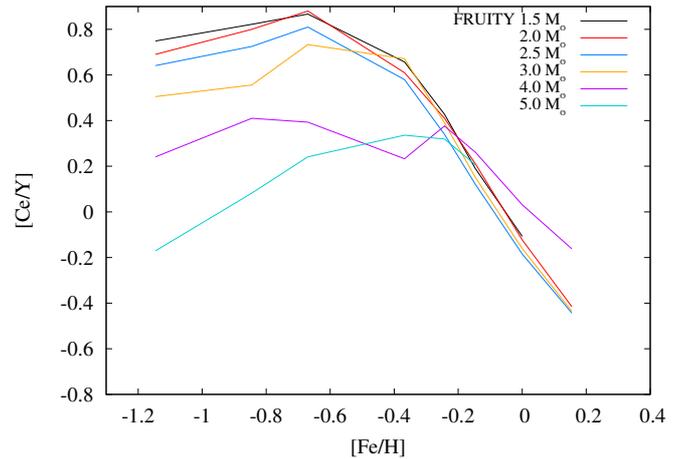}
      \caption{The [Ce/Y] ratio versus [Fe/H] at the stellar surface at the end of the evolution from all the FRUITY models with [s/Fe]$>$0.25 reported in Table~\ref{tab:FRUITY}. The different colours represent different stellar masses, as indicated. \label{fig:FRUITY}}
      \end{figure}

\begin{figure}
   \centering
   \includegraphics[width=\hsize]{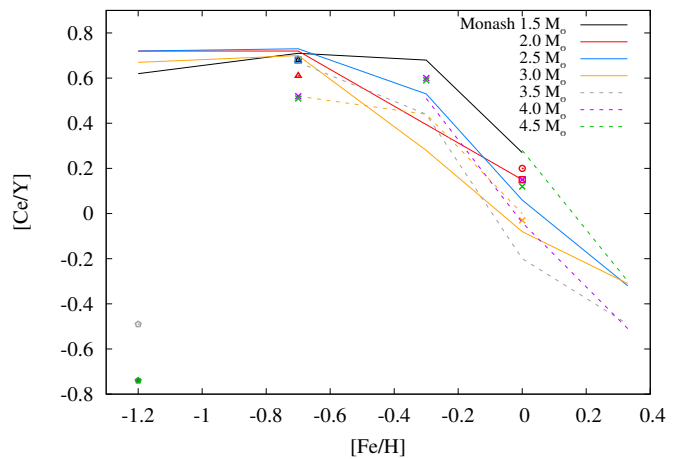}
      \caption{The [Ce/Y] ratio versus [Fe/H] at the stellar surface at the end of the evolution from all the Monash models with [s/Fe]$>$0.25 reported in Table~\ref{tab:monash}. As in Fig.~\ref{fig:FRUITY}, the different colours represent different stellar masses. Furthermore, the solid lines represent the models with $M_{\rm mix}=2 \times 10^{-3}$ \(M_\odot\), the dashed lines with $M_{\rm mix}=1 \times 10^{-3}$ \(M_\odot\), and the different types of dots represent the following $M_{\rm mix}$ values:
      circle = $1 \times 10^{-3}$ \(M_\odot\); square = $4 \times 10^{-3}$ \(M_\odot\); triangle = $6 \times 10^{-3}$ \(M_\odot\); cross = $1 \times 10^{-4}$ \(M_\odot\); diamond = 0. \label{fig:monash}}
      \end{figure}

The details of the models have been reported elsewhere and we refer the reader to the papers mentioned above. All the models presented here were calculated using initial solar scaled abundances from \citet{asplund09}. Some FRUITY models were further computed including stellar rotation \citep{piersanti13} and we will include some of these models in the discussion.
In Table~\ref{tab:FRUITY} and Fig.~\ref{fig:FRUITY} we present the results from the FRUITY models and in Table~\ref{tab:monash} and Fig.~\ref{fig:monash} those from the Monash models. We select only models that are able to achieve [s/Fe]$>$0.25 dex at the stellar surface, since this is the same condition applied by deC16 to define Ba stars. However, 
the Ba stars from deC16 are all giants and dilution of the $s$-process AGB material deposited on the stellar surface necessarily occurred due to the presence of their convective envelope. Consequently, only the models that achieve [s/Fe] above roughly 0.5 dex are appropriate for the comparison here, as will be discussed in detail at the start of Sec.~\ref{ref:results}. On the other hand, all the models we present here are appropriate for future comparison to observations of Ba dwarfs. Even if these are less luminous and more difficult to observe than the Ba giants considered here, the number of known Ba dwarfs is continuously growing \citep[see, e.g.,][and references therein]{kong18}. In the case of Ba dwarfs, the material deposited on the stellar surface during the mass transfer may remain undiluted, although several mixing processes could occur that may carry it deeper into the star and dilute it. These processes have been investigated in detail mostly in relation to CEMP stars \citep[see, e.g.,][and references therein]{matrozis17a}.

The FRUITY models include the formation of the \iso{13}C pocket in a self-consistent way based on a time-dependent overshoot mechanisms. The free overshoot parameter $\beta$ in the exponentially decaying velocity function is set such as enough neutrons are released in order to maximize the production of $s$-process elements (see \citealt{cristallo09} for details).
On the other hand, in the Monash models the mixing of protons leads to the formation of \iso{13}C pockets is performed by adding an exponential profile of the proton abundance into the top layers of the He-intershell over an extent in mass $M_{\rm mix}$, which is a free parameter. A detailed description of this method and related equations is given in \citet{buntain17}. These authors also showed that varying the steepness of the exponential function or the variable $M_{\rm mix}$ produces similar results in terms of the $s$-process abundance distribution. Here, we consider models where only $M_{\rm mix}$ is varied and show them in Table~\ref{tab:monash}. A few of the Monash models with the lowest mass or the highest metallicity were calculated including non-time-dependent overshoot by extending the base of the envelope by $N_{\rm ov}$ (the value of which is indicated in the footnotes of Table~\ref{tab:monash}) pressure scale heights during the TDU. This overshoot does not lead to the partial mixing needed to produce the \iso{13}C pocket, but has the effect of enhancing the TDU efficiency and produce C stars of low initial masses, as required by observations \citep[see discussion in][]{kamath12,karakas16}.

Careful comparisons between the two sets of models in relation to the $s$-process results can be found in \citet{fishlock14} and \citet{karakas16}. 
The main difference between the FRUITY and the Monash sets of models are the absolute values of the elemental abundances, i.e., the [X/Fe] ratios. The FRUITY models have typically lower abundances than the Monash models,  depending on the size of $M_{\rm mix}$. For example, in the case of the 3 \msun\ model of Z=0.014, the Monash model with $M_{\rm mix}=2 \times 10^{-3}$ \msun, results in [Ce/Fe] roughly 0.5 dex higher than the corresponding FRUITY model. This has been noted before and is due also to a different efficiency of the TDU. 
As demonstrated by Tables~\ref{tab:FRUITY} and \ref{tab:monash}, the elements belonging to the first peak (Y and Zr) behave very similarly to each other, with [X/Fe] values mostly within 0.15 dex. The same applies to the elements belonging to the second peak (Ce and Nd), although some differences of up to 0.3 dex are present in this case in the Monash models of higher mass.

\begin{table*}
\caption{FRUITY models that achieve [s/Fe]$>$0.25 at the stellar surface in the metallicity range relevant to Ba stars, computed without rotation. The mass (in \(M_\odot\)), metallicity, total number of TDU episodes (TDU$_{\rm tot}$), final surface abundances [X/Fe] of selected elements and the [Ce/Y] ratios are indicated. \label{tab:FRUITY}}
\centering
\begin{tabular}{lcccccccccc}
M(\(M_\odot\)) & Z & TDU$_{\rm tot}$ & [Y/Fe] & [Zr/Fe] & [Rb/Fe] & [La/Fe] & [Ce/Fe] & [Nd/Fe] & [Pb/Fe] & [Ce/Y]\\
\hline\hline
1.5	&	0.001	&	8	&	0.711	&	0.739	&	0.254	&	1.35	&	1.46	&	1.36	&	2.46	&	0.749	\\
	&	0.002	&	7	&	0.798	&	0.858	&	0.297	&	1.50	&	1.62	&	1.53	&	2.38	&	0.822	\\
	&	0.003	&	6	&	0.713	&	0.786	&	0.198	&	1.47	&	1.58	&	1.48	&	1.93	&	0.867	\\
	&	0.006	&	7	&	0.813	&	0.875	&	0.193	&	1.42	&	1.47	&	1.33	&	1.17	&	0.657	\\
	&	0.008	&	5	&	0.682	&	0.734	&	0.145	&	1.09	&	1.11	&	0.96	&	0.642	&	0.428	\\
	&	0.01	&	4	&	0.715	&	0.748	&	0.207	&	0.913	&	0.904	&	0.742	&	0.399	&	0.189	\\
	&	0.014	&	5	&	0.621	&	0.62	&	0.148	&	0.586	&	0.514	&	0.368	&	0.152	&	$-$0.107	\\
\hline				 	   			                                                           		
2.0	&	0.001	&	11	&	1.05	&	1.08	&	0.701	&	1.63	&	1.74	&	1.63	&	2.65	&	0.69	\\
	&	0.002	&	11	&	1.04	&	1.08	&	0.465	&	1.72	&	1.84	&	1.74	&	2.55	&	0.80    \\
	&	0.003	&	11	&	1.05	&	1.11	&	0.395	&	1.82	&	1.93	&	1.82	&	2.19	&	0.88	\\
	&	0.006	&	10	&	1.15	&	1.21	&	0.389	&	1.72	&	1.76	&	1.61	&	1.47	&	0.61	\\
	&	0.008	&	10	&	1.11	&	1.15	&	0.338	&	1.50	&	1.52	&	1.36	&	1.11	&	0.41	\\
	&	0.01	&	9	&	1.11	&	1.13	&	0.362	&	1.36	&	1.32	&	1.13	&	0.857	&	0.21	\\
	&	0.014	&	9	&	1.01	&	1.00	&	0.305	&	0.957	&	0.887	&	0.707	&	0.496	&	$-$0.123	\\
	&	0.02	&	9	&	0.923	&	0.865	&	0.306	&	0.56	&	0.507	&	0.375	&	0.189	&	$-$0.416	\\
\hline										                                                        		
2.5	&	0.001	&	10	&	0.939	&	0.995	&	0.676	&	1.48	&	1.58	&	1.48	&	2.53	&	0.641	\\
	&	0.002	&	10	&	0.925	&	0.957	&	0.526	&	1.53	&	1.65	&	1.53	&	2.45	&	0.725	\\
	&	0.003	&	11	&	1.03	&	1.06	&	0.462	&	1.73	&	1.84	&	1.73	&	2.13	&	0.81	\\
	&	0.006	&	13	&	1.27	&	1.32	&	0.560	&	1.80	&	1.85	&	1.71	&	1.57	&	0.58	\\
	&	0.008	&	15	&	1.30	&	1.33	&	0.505	&	1.63	&	1.64	&	1.47	&	1.26	&	0.34	\\
	&	0.01	&	14	&	1.28	&	1.30	&	0.486	&	1.43	&	1.40	&	1.22	&	1.00	&	0.12	\\
	&	0.014	&	13	&	1.18	&	1.16	&	0.466	&	1.03	&	0.995	&	0.819	&	0.621	&	$-$0.185  \\
	&	0.02	&	15	&	1.07	&	0.999	&	0.413	&	0.643	&	0.626	&	0.485	&	0.256	&	$-$0.444	\\
\hline				 	   			                                                            		
3.0	&	0.001	&	11	&	0.885	&	0.926	&	0.776	&	1.33	&	1.39	&	1.32	&	2.39	&	0.505	\\
	&	0.002	&	10	&	0.844	&	0.871	&	0.623	&	1.32	&	1.40	&	1.31	&	2.29	&	0.556	\\
	&	0.003	&	9	&	0.697	&	0.732	&	0.364	&	1.32	&	1.43	&	1.33	&	2.03	&	0.733	\\
	&	0.006	&	9	&	0.849	&	0.895	&	0.312	&	1.46	&	1.52	&	1.38	&	1.29	&	0.671	\\
	&	0.008	&	10	&	1.03	&	1.06	&	0.369	&	1.41	&	1.42	&	1.26	&	1.04	&	0.39	\\
	&	0.01	&	11	&	1.05	&	1.06	&	0.361	&	1.24	&	1.20	&	1.02	&	0.796	&	0.15	\\
	&	0.014	&	13	&	1.06	&	1.04	&	0.375	&	0.965	&	0.899	&	0.726	&	0.539	&	$-$0.161	\\
	&	0.02	&	14	&	1.01	&	0.931	&	0.380	&	0.615	&	0.574	&	0.439	&	0.235	&	$-$0.436	\\
\hline				 	   			                                                            		
4.0	&	0.001	&	15	&	0.617	&	0.612	&	0.709	&	0.838	&	0.858	&	0.801	&	1.83	&	0.241	\\
	&	0.002	&	15	&	0.462	&	0.479	&	0.441	&	0.827	&	0.872	&	0.804	&	1.75	&	0.41	\\
	&	0.003	&	12	&	0.507	&	0.511	&	0.368	&	0.846	&	0.90	&	0.807	&	1.59	&	0.393	\\
	&	0.006	&	9	&	0.728	&	0.693	&	0.402	&	0.913	&	0.961	&	0.859	&	1.04	&	0.233	\\
	&	0.008	&	9	&	0.405	&	0.41	&	0.171	&	0.745	&	0.781	&	0.684	&	0.663	&	0.376	\\
	&	0.01	&	8	&	0.436	&	0.44	&	0.159	&	0.691	&	0.697	&	0.59	&	0.468	&	0.261	\\
	&	0.014	&	8	&	0.42	&	0.414	&	0.126	&	0.478	&	0.451	&	0.351	&	0.231	&	0.031	\\
	&	0.02	&	8	&	0.569	&	0.536	&	0.168	&	0.43	&	0.407	&	0.312	&	0.189	&	$-$0.162	\\
\hline																                            		
5.0	&	0.001	&	24	&	0.527	&	0.469	&	0.812	&	0.378	&	0.356	&	0.326	&	1.23	&	$-$0.171	\\
	&	0.002	&	22	&	0.351	&	0.313	&	0.570	&	0.417	&	0.433	&	0.394	&	1.24	&	0.082	\\
	&	0.003	&	18	&	0.26	&	0.256	&	0.286	&	0.466	&	0.501	&	0.439	&	1.18	&	0.241	\\
	&	0.006	&	12	&	0.227	&	0.23	&	0.144	&	0.511	&	0.563	&	0.486	&	0.739	&	0.336	\\
	&	0.008	&	11	&	0.27	&	0.274	&	0.150	&	0.554	&	0.589	&	0.501	&	0.522	&	0.319	\\
	&	0.01	&	11	&	0.215	&	0.217	&	0.0988	&	0.414	&	0.421	&	0.344	&	0.242	&	0.206	\\

\hline
\end{tabular}
\end{table*}

\begin{table*}
\caption{Same as Table~\ref{tab:FRUITY} but for the Monash models. In this table a further column (3) is added, which indicates the value of $M_{\rm mix}$ in units of \(M_\odot\). \label{tab:monash}}
\begin{center}
\begin{tabular}{lccccccccccc}
M & Z & $M_{\rm mix}$ & TDU$_{\rm tot}$ & [Y/Fe] & [Zr/Fe] & [Rb/Fe] & [La/Fe] & [Ce/Fe] & [Nd/Fe] & [Pb/Fe] & [Ce/Y]\\
\hline\hline

1.5	&	0.001	& $2 \times 10^{-3}$	&	10	&	1.099	&	1.179	&	0.520	&	1.617	&	1.723	&	1.625	&	2.681	&	0.625		\\
	&	0.0028	&	$2 \times 10^{-3}$	  &	7		&	0.922	&	0.960	&	0.291	&	1.516	&	1.633	&	1.538	&	2.200	&	0.711		\\
	&			&	$6 \times 10^{-3}$	  &	7		&	1.319	&	1.377	&	0.553	&	1.910	&	1.998	&	1.891	&	2.428	&	0.679		\\
	&	0.007$^a$	&	$2 \times 10^{-3}$ &	5		&	0.730	&	0.781	&	0.152	&	1.293	&	1.381	&	1.274	&	1.367	&	0.652		\\
	&	0.014$^b$	&	$2 \times 10^{-3}$ &	4		&	0.684	&	0.728	&	0.163	&	0.962	&	0.947	&	0.797	&	0.424	&	0.263		\\
\hline	            	                                    	        														
2.0	&	0.001	&	$2 \times 10^{-3}$	  &	14	&	1.418	&	1.526	&	0.697	&	2.031	&	2.139	&	2.022	&	2.954	&	0.721		\\
	&	0.0028	&	$2 \times 10^{-3}$	  &	13		&	1.401	&	1.435	&	0.701	&	2.001	&	2.118	&	2.014	&	2.678	&	0.717		\\
	&			&	$6 \times 10^{-3}$	  &	13		&	1.827	&	1.881	&	1.119	&	2.371	&	2.436	&	2.318	&	2.729	&	0.609		\\
	&	0.014	&	$1 \times 10^{-3}$	  &	8		&	0.889	&	0.948	&	0.193	&	1.112	&	1.087	&	0.917	&	0.677	&	0.198		\\
	&			&	$2 \times 10^{-3}$	  &	8		&	1.155	&	1.199	&	0.337	&	1.308	&	1.299	&	1.135	&	1.055	&	0.144		\\
	&			&	$4 \times 10^{-3}$	  &	8		&	1.361	&	1.388	&	0.497	&	1.492	&	1.514	&	1.368	&	1.402	&	0.152		\\
\hline	            	                                                														
2.5	&	0.001	&	$2 \times 10^{-3}$	  &	16	&	1.544	&	1.677	&	1.199	&	2.134	&	2.265	&	2.072	&	2.940	&	0.721		\\
	&	0.0028	&	$2 \times 10^{-3}$	  &	17		&	1.633	&	1.681	&	1.026	&	2.253	&	2.365	&	2.239	&	2.596	&	0.731		\\
	&			&	$4 \times 10^{-3}$	  &	17		&	1.862	&	1.913	&	1.400	&	2.433	&	2.510	&	2.376	&	2.726	&	0.648		\\
	&	0.007	&	$2 \times 10^{-3}$	  &	15		&	1.499	&	1.560	&	0.668	&	1.983	&	2.030	&	1.888	&	1.896	&	0.531		\\
	&	0.014	&	$2 \times 10^{-3}$	  &	12		&	1.355	&	1.377	&	0.533	&	1.414	&	1.411	&	1.255	&	1.224	&	0.057		\\
	&	0.03$^c$	&	$2 \times 10^{-3}$ &	12		&	1.223	&	1.177	&	0.480	&	0.927	&	0.905	&	0.747	&	0.510	&	$-$0.319	\\
\hline	                                                                														
3.0	&	0.001	&	$5 \times 10^{-4}$ &	20	&	1.082	&	1.272	&	1.002	&	1.594	&	1.754	&	1.532	&	2.563	&	0.671\\
	&	0.0028	&	$1 \times 10^{-3}$	  &	17	&	1.467	&	1.507	&	1.204	&	1.861	&	1.992	&	1.799	&	2.246	&	0.525		\\
	&			&	$2 \times 10^{-3}$	  &	17	&	1.521	&	1.580	&	1.453	&	2.095	&	2.223	&	2.016	&	2.602	&	0.703		\\
	&	0.007	&	$1 \times 10^{-3}$	  &	19  	&	1.414	&	1.463	&	0.751	&	1.819	&	1.850	&	1.691	&	1.567	&	0.437		\\
	&			&	$2 \times 10^{-3}$	  &	19	&	1.680	&	1.721	&	1.038	&	1.954	&	1.965	&	1.792	&	1.775	&	0.284		\\
	&	0.014	&	$1 \times 10^{-4}$	  &	17		&	0.412	&	0.431	&	0.076	&	0.430	&	0.376	&	0.247	&	0.042	&	$-$0.036	\\
	&			&	$1 \times 10^{-3}$	  &	17		&	1.310	&	1.338	&	0.542	&	1.343	&	1.305	&	1.126	&	0.931	&	$-$0.005	\\
	&			&	$2 \times 10^{-3}$	  &	17		&	1.537	&	1.527	&	0.798	&	1.457	&	1.455	&	1.297	&	1.198	&	$-$0.082	\\
	&	0.03$^a$	&	$2 \times 10^{-3}$ &	16		&	1.227	&	1.152	&	0.584	&	0.936	&	0.919	&	0.767	&	0.557	&	$-$0.308	\\
\hline	            	                                                														
3.5	&	0.001	&	0	&	27	&	0.577	&	0.514	&	0.890	&	0.096	&	0.087	&	0.018	&	0.043	&	$-$0.491	\\
	&	0.0028	&	$1 \times 10^{-3}$	&	21	&	1.302	&	1.396	&	1.341	&	1.825	&	1.972	&	1.761	&	2.398	&	0.671		\\
	&	0.007	&	$1 \times 10^{-3}$	&	19	&	1.412	&	1.478	&	1.136	&	1.798	&	1.852	&	1.607	&	1.543	&	0.439		\\
	&	0.014	&	$1 \times 10^{-3}$	&	19		&	1.344	&	1.331	&	0.777	&	1.172	&	1.141	&	0.965	&	0.801	&	$-$0.203	\\
	&	0.03	&	$1 \times 10^{-3}$	&	24		&	1.103	&	1.003	&	0.598	&	0.638	&	0.607	&	0.467	&	0.255	&	$-$0.496	\\
\hline	            		            			                    														
4.0	&	0.001	&	0	&	68	&	1.417	&	1.396	&	1.650	&	0.723	&	0.680	&	0.445	&	0.294	&	$-$0.738	\\
	&	0.0028	&	$1 \times 10^{-4}$	&	24	&	0.540	&	0.601	&	0.607	&	0.924	&	1.063	&	0.870	&	1.736	&	0.523		\\
	&	0.007	&	$1 \times 10^{-4}$	&	23	&	0.561	&	0.630	&	0.466	&	1.030	&	1.160	&	0.953	&	1.073	&	0.599		\\
	&			&	$1 \times 10^{-3}$	&	23	&	1.349	&	1.428	&	1.340	&	1.758	&	1.861	&	1.618	&	1.974	&	0.512		\\
	&	0.014	&	$1 \times 10^{-4}$	&	20		&	0.513	&	0.540	&	0.276	&	0.676	&	0.656	&	0.469	&	0.165	&	0.143		\\
	&			&	$1 \times 10^{-3}$	&	20		&	1.280	&	1.278	&	1.120	&	1.223	&	1.238	&	0.997	&	0.950	&	$-$0.042	\\
	&	0.03	&	$1 \times 10^{-3}$	&	20		&	1.036	&	0.918	&	0.713	&	0.572	&	0.535	&	0.389	&	0.175	&	$-$0.501	\\
\hline	            		            			                    														
4.5	&	0.001	&	0	&	78	&	1.425	&	1.410	&	1.646	&	0.729	&	0.690	&	0.455	&	0.305	&	$-$0.735	\\
	&	0.0028	&	$1 \times 10^{-4}$	&	30		&	0.610	&	0.668	&	0.712	&	0.982	&	1.124	&	0.934	&	1.808	&	0.515		\\
	&	0.007	&	$1 \times 10^{-4}$	&	50		&	0.780	&	0.867	&	0.848	&	1.235	&	1.366	&	1.139	&	1.188	&	0.586		\\
	&	0.014	&	$1 \times 10^{-4}$	&	29		&	0.629	&	0.682	&	0.435	&	0.765	&	0.750	&	0.525	&	0.223	&	0.120		\\
	&			&	$1 \times 10^{-3}$	&	29		&	1.314	&	1.350	&	1.325	&	1.514	&	1.590	&	1.331	&	1.685	&	0.275		\\
	&	0.03	&	$1 \times 10^{-3}$	&	16		&	0.855	&	0.768	&	0.793	&	0.586	&	0.559	&	0.400	&	0.248	& $-$0.296      \\
\hline
\end{tabular}
\end{center}
\footnotesize a: $N_{\rm ov}$ $=$ 1.0, \footnotesize b: $N_{\rm ov}$ $=$ 3.0, \footnotesize c: $N_{\rm ov}$ $=$ 2.5\\
\end{table*}

The most obvious feature of both sets of models in Tables~\ref{tab:FRUITY} and \ref{tab:monash} is the shift of the peak abundance production from the first to the second and to the third $s$-process peak as the metallicity decreases. Figures~\ref{fig:FRUITY} and \ref{fig:monash} show the [Ce/Y] ratios as function of metallicity for all the models considered here. All the models up to 3 \msun\ where the \iso{13}C is the main neutron source show the same trend with the [Ce/Y] ratios increasing with decreasing the [Fe/H]. This is a well known feature of the \iso{13}C neutron source \citep{busso01} and a fundamental consequence of the fact that the neutron exposure $\tau$ in the \iso{13}C pocket is proportional to the \iso{13}C/\iso{56}Fe abundances ratio. This ratio increases with decreasing the metallicity because the number of Fe seeds decreases with metallicity, while the number of \iso{13}C nuclei does not change with metallicity. The \iso{13}C nuclei are primary (i.e., metallicity independent) because they are produced by the interaction of H with the \iso{12}C produced by He burning \citep{clayton88}. In other words, fewer Fe seeds capturing neutrons means that more free neutrons are available for progressing to the production of the heaviest elements. The maximum [Ce/Y] value is around 0.8 in both sets of models, however, at higher metallicities the [Ce/Y] ratios from the Monash models are typically higher, at most by roughly 0.3 dex, than the FRUITY models. This may be partly due to the details of the different implementation for the formation of the \iso{13}C pocket, where \citet{buntain17} found differences of roughly 0.1 dex for proton profiles not drastically different from each other (see 1.8 \msun\ models in their Table 5), as well as lower temperatures in the intershell in the FRUITY models leading to incomplete burning of \iso{13}C before the onset of the next TP. 

The models by \citet{ritter17} do not produce high enough Ba and $s$-process abundances to be able to explain the level of enrichment observed in the metallicity range of Ba stars in the deC16 sample. This is due to the prescription used for the convective-boundary mixing at the bottom of the convective envelope during the TDU, with an exponential-diffusive model based on \citet{freytag:96} and \citet{herwig:00}. This results in \iso{13}C pockets smaller and less efficient in making $s$-process elements compared to the other models considered here. In the AGB models by \citet{battino16}, the efficiency of \iso{13}C pocket formation is higher with respect to the models by \citet{ritter17} due to the inclusion of the effect of gravity waves, as according to \citet{denissenkov:03}. These models produce [s/Fe] in the range observed in Ba stars, from 0.5 to 1.5 dex, and [hs/ls] ratios from 0 to 0.5 dex for stars of around solar metallicity, within the same range predicted by the Monash and the FRUITY models.  
Note that both the models by Ritter et al. and by Battino et al. employ convective-boundary mixing also at the bottom of convective TPs, based on the study of \citet{herwig:07}. This results in an increased \iso{12}C abundance in the He intershell, a higher \iso{13}C abundance in the \iso{13}C pocket, and a local higher number of neutron captures per Fe seeds \citep{lugaro03}. 

As shown in Figs.~\ref{fig:FRUITY} and \ref{fig:monash}, below [Fe/H] $\sim$ $-$0.6 the [Ce/Y] ratio becomes flat as equilibrium is achieved between the first two peaks while the $s$-process flux reaches the third peak at Pb. 
The production of Pb located at the end of the $s$-process chain of neutron captures increases steadily with decreasing metallicity \citep[][]{Gallino98,vaneck01}, with the result that most of the cosmic Pb is made in low-metallicity AGB stars \citep{travaglio01}.

The effect of the \iso{22}Ne neutron source becomes more pronounced with increasing stellar mass.
Similarly to the \iso{13}C reaction, this neutron source also becomes more efficient in producing heavier elements as the metallicity decreases. The \iso{22}Ne abundance derives from the initial CNO abundance in the star and thus it decreases with the metallicity together with the Fe seeds. However, there is a primary component to \iso{22}Ne due to the TDU of primary \iso{12}C from partial He burning - \iso{12}C is converted in \iso{14}N via H burning, which, in turn, produces \iso{22}Ne via double $\alpha$-capture during He burning. 
The \iso{22}Ne neutron source is activated efficiently in AGB stars of initial mass above ~4--5 \msun, the exact range depending on the metallicity, where it contributes to the production of the bulk of the elements heavier than Fe. In these models, we still see enhancements in the elements heavier than Fe even if the \iso{13}C($\alpha$,n)\iso{16}O reaction is not present (see, e.g., the Monash models of masses between 3.5 and 4.5 \msun\ at Z=0.001 with $M_{\rm mix}=0$). In stars of mass between ~2.5 \msun\ and ~4 \msun, the exact range again depending on the metallicity, the \iso{22}Ne neutron source is also activated, but marginally. In this case it does not significantly contribute to the production of the bulk of the elements heavier than Fe, but it can act upon and affect the distribution produced by the \iso{13}C source. In these models if the \iso{13}C source is not present, we do not have any significant production of $s$-process elements. For example, the Monash 3.75 \msun\ model at Z=0.007 with $M_{\rm mix}=0$ produces only [Sr/Fe]=0.024 and [Ba/Fe]=0.015 \citep[see][]{karakas16}.

The activation of the \iso{22}Ne source in our models of higher mass ($>$2.5 \msun) results in higher production of the first peak $s$-process elements for metallicities around solar, and higher production of the second peak $s$-process elements for lower metallicities.  This is due to the neutron exposure produced by the \iso{22}Ne source
being roughly an order of magnitude lower than that produced by the \iso{13}C.  This effect is more noticeable in the Monash than in the FRUITY models. In fact, at the earlier TDUs, the Monash models of masses above roughly 2.5 \msun\ produce similar [Ce/Y] ratios as the FRUITY models. Later on, when the temperature increases and the \iso{22}Ne source is more significantly activated, the [Ce/Y] ratios decrease. For example, the Monash 3 \msun, [Fe/H]=$-$0.3 model produces [Ce/Y]$\sim$0.6 after the first few TDUs, very similar to the FRUITY model, however, as the TDU number increases the ratio decreases to the final values around 0.3 dex seen in the Fig.~\ref{fig:monash}.  
Another effect of the \iso{22}Ne neutron source noticeable in Tables~\ref{tab:FRUITY} and \ref{tab:monash} is the production of Rb \citep{abia01}. This depends on the activation of the branching point at \iso{86}Rb \citep{vanraai12} for the higher neutron densities associated to the \iso{22}Ne neutron source (up to $10^{13}$ cm$^3$) with respect to the \iso{13}C source (up to $10^{8}$ cm$^3$). In fact, for the low mass stars [Rb/Fe]$<$[Zr/Fe], while for the high mass stars the [Rb/Fe] increases and the reverse applies in some cases. 
Further differences appear between the FRUITY and the Monash predictions for AGB stars of masses above roughly 4 \msun. In the Monash models these masses show a similar behaviour to the lower masses, while in the FRUITY models the 4 and 5 \msun\ stars deviate from the trend of the lower masses in that the [Ce/Y] ratios remains below 0.4.

Finally, we note that in the Monash models of mass above 3 and metallicity 0.001 \citep{fishlock14} the \iso{13}C pocket was not included and the resulting $s$-process distribution is dominated by the effect of the \iso{22}Ne source producing much more favourably the first rather than the second $s$-process peak. The resulting [Ce/Y] ratios are negative. There are only three Ba stars at this metallicity and they show positive [Ce/Y], which appears to exclude these models as a fit for these stars. 

\section{Results and discussion}
\label{ref:results}

   \begin{figure}
   \centering
   \includegraphics[width=\hsize]{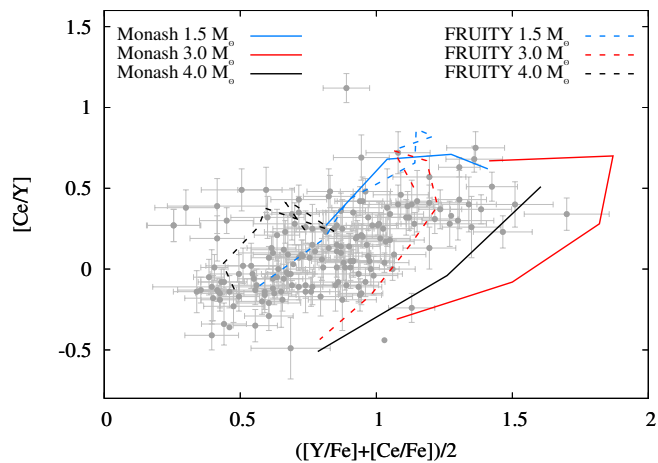}
      \caption{The [Ce/Y] ratios plotted against the average enhancements of the two elements for the Ba stars and the final surface composition a selection of models with the indicated stellar masses. For the 1.5 and 3 \msun\ Monash models we used the cases with $M_{\rm mix}=2 \times 10^{-3}$ \msun, for the 4 \msun\ Monash models the cases with $M_{\rm mix}=1 \times 10^{-3}$ \msun. The dots without error bars represent stars for which there are less than 3 lines for one of the elements.
         \label{fig:cefe}}
   \end{figure}

In Fig.~\ref{fig:cefe} we plot the [Ce/Y] ratio as function of the overall enhancement of the two elements. In this and the following figures we selected from our full set of models those of mass 1.5, 3, and 4 \msun. This range well represents the observations of the masses of Ba stars, which show a peak at around 2.5 \msun\ with a spread around it of roughly $\pm$1.5 \msun \citep{deC,escorza17}. The rest of the models shown in Figs.~\ref{fig:FRUITY} and \ref{fig:monash} are within the values of those plotted in the following figures. 
Overall, both the data and the models show a qualitatively similar trend:  higher [Ce/Y] ratio are expected for higher $s$-process enhancements. This is a typical feature of the $s$ process because higher neutron exposures naturally result in an increase of both the absolute amount of abundances produced (represented by the x axis in the figure) and a shift towards the second $s$-process peak (represented by the y axis). 
Any [X/Fe] ratio, where X is a generic $s$-process element or a combination of $s$-process elements such as in the x axis of Fig.~\ref{fig:cefe}, is affected by the binary transfer and accretion mechanism not considered in our models that determines the amount of $s$-process material carried from the primary to the secondary star. This is because Fe is not significantly affected by AGB nucleosynthesis. The accretion mechanism controls which fraction of the total matter lost by the primary star is deposited onto the secondary. 
Furthermore, if the Ba star is a giant, as in the case of all the stars considered here, the material deposited at its surface is mixed with the whole stellar envelope and further diluted. 

Figure~\ref{fig:cefe} shows that the 3 \msun\ models allow for an overall dilution factor between 0.5 and 1.5 dex. This dilution factor corresponds to a horizontal shift of the model predictions. If we assume for simplicity that the accreted material is mixed with an envelope mass of roughly 2 \msun, this dilution factor translates into roughly 0.7 to 0.07 \msun\ that need to be accreted from the primary star. The 1.5 \msun\ models instead only allow for a dilution factor around 0.5 dex, which translates in 0.3 \msun\ of accreted mass, considering this time an envelope of 1 \msun.
The 4 \msun\ Monash models computed with $M_{\rm mix} = 1 \times 10^{-4}$ \msun\ produces a similar result as the 3 \msun\ models, while only at low metallicity the FRUITY models of the same mass allow for dilution due to mass accretion. The predicted [X/Fe] values depend on the efficiency of the TDU, the mass loss, and the extent of the \iso{13}C pocket. These are the three major AGB model uncertainties, as a consequence the models cannot set strong constraints on the accretion process. 

One the other hand, if we consider [X/Y] ratios where both elements are produced by the $s$ process, the dilution due to mass transfer and mixing on the secondary applies to both the elements and in first approximation it is factored out when taking the ratios. For example, the [hs/ls] ratio has been extensively used as a direct measure of the neutron exposure $\tau$ in the intershell of AGB stars. This is because the relative accumulation of the $s$-process peaks is a strong function of the total number of available neutrons, represented by the total time-integrated neutron flux, or $\tau$. 

   \begin{figure*}
   \centering
   \includegraphics[scale=1.5]{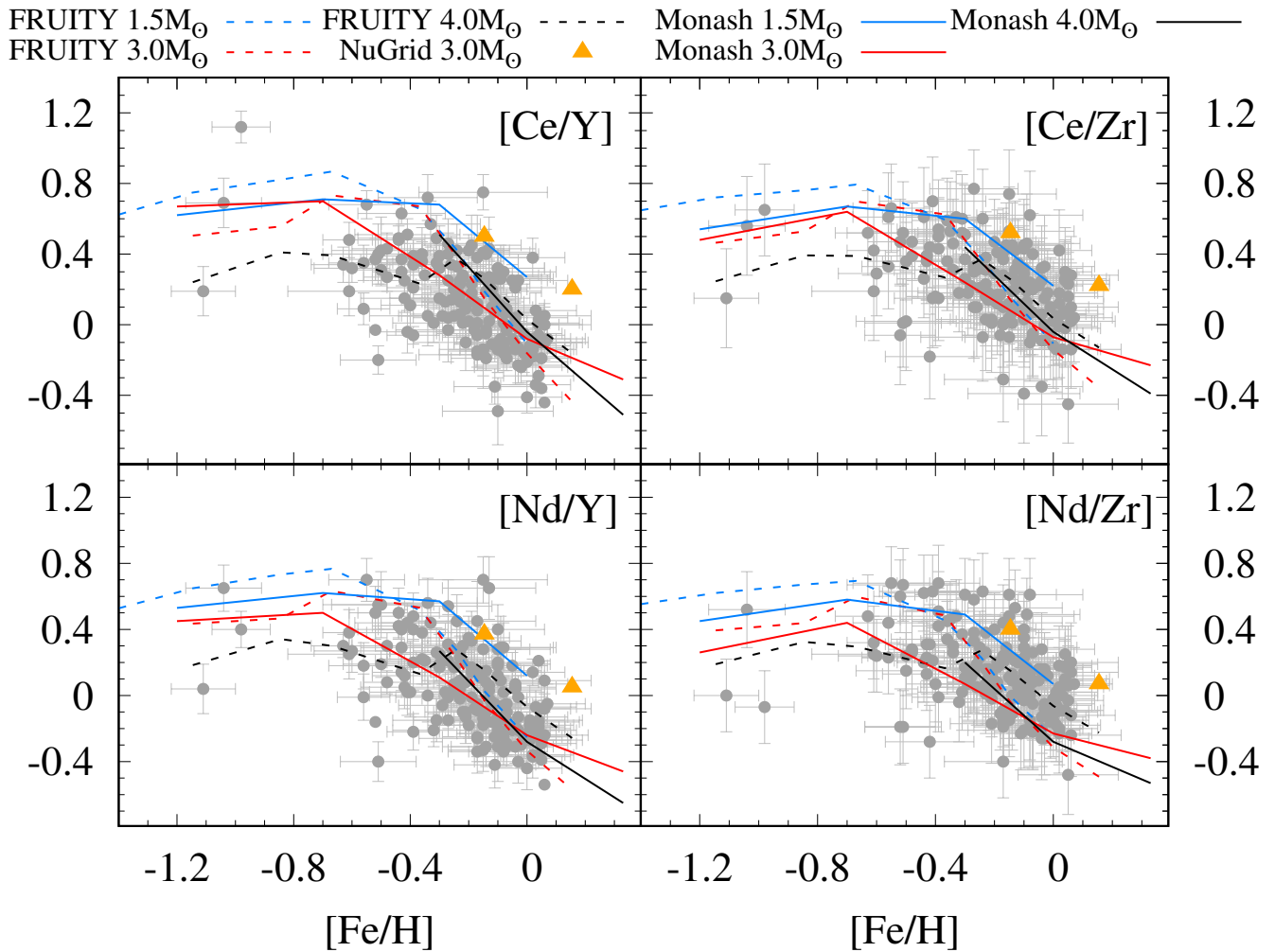}
      \caption{Comparison between Ba star observations and the predicted final surface composition for the same selection of FRUITY and Monash models as in Fig.~\ref{fig:cefe}. We show also 3 \msun\  type He07 models from \citet{battino16} for comparison. We consider all the four combinations of the ls (Y, Zr) and the hs (Ce, and Nd) elements. The dots without error bars represent stars for which there are less than 3 lines for one of the elements.\label{fig:multi}}
   \end{figure*}

In Fig.~\ref{fig:multi} we combine the different possibilities of ratios between the first and the second peak $s$-process elements observed in the Ba stars by deC16 and compare them to the stellar models. We note that while the predicted [X/Fe] are calculated normalising to the solar meteoritic abundances from \citet{asplund09}, the observations are normalised to the solar photospheric abundances given in \citet{g_s98}. For the species taken into consideration in this paper the largest difference is of 0.07 dex for both Y and Zr, which is comparable to typical uncertainties in the solar abundances themselves. Based on our error calculations the ratios computed using Zr show higher error bars than those computed using Y. The plots show the predicted final surface composition of the AGB stars, i.e., after the last computed TDU episode, while the mass transfer could have occurred earlier. The plotted ratios, however, are not significantly different at earlier times, as far as enough $s$-process abundances are present at the stellar surface to allow for dilution into the Ba star envelope. The main result is that the data trend of the $s$-process ratios increasing with decreasing [Fe/H] between 0 and $-$0.8 is matched by the theoretical trend of the models. This clearly confirms the primary behaviour of the main \iso{13}C neutron source. 

Several second-order effects can produce the spread of roughly a factor of three at any given [Fe/H] (see also Fig.~\ref{fig:cey}): variations in the initial mass, which can affect the activation of the \iso{22}Ne neutron source and the temperature in the \iso{13}C pocket; the treatment of the mixing at the base of the TP, as well as during the TDU (see, e.g., the difference between the "Standard" and the "Tail" FRUITY models in \citealt{cristallo15b} and the magnetic models of \citealt{trippella16}), and/or mixing within the \iso{13}C pocket possibly due to stellar rotation and/or magnetic fields. 
Due to the uncertainties associated with all these processes it is not possible yet to accurately establish among these possibilities the actual physics from which the spread originates. Furthermore, the exact location of the theoretical spread cannot be firmly established until systematic uncertainties in the neutron-capture cross sections of the nuclei involved are resolved. For example, the neutron-capture cross section of the main $s$-process seed nucleus \iso{56}Fe is uncertain at the temperature of 90 MK of the activation of the \iso{13}C neutron source because experimental data are currently available only for temperature around 270 MK and the lower-temperature values are derived via extrapolation using theoretical models. As an exercise, we calculated a 3 \msun, Z=0.014 Monash model increasing the neutron-capture cross section of \iso{56}Fe by 50\% in the whole temperature
range. The result is a decrease of the [Ce/Y] ratio of
roughly 10\% (i.e., 0.04 dex). The decrease is expected since if \iso{56}Fe captures more neutrons, the neutron exposure decreases, however, the effect is not linear. Furthermore, new evaluations of the neutron-capture cross sections of \iso{140}Ce and \iso{89}Y, of crucial importance here, are currently undergoing at the n\_TOF experiment at CERN \citep{amaducci18,tagliente17}. The new values may result in a systematic shift of the plotted lines.

\subsection{The case of stellar rotation}

   \begin{figure}
   \centering
   \includegraphics[width=\hsize]{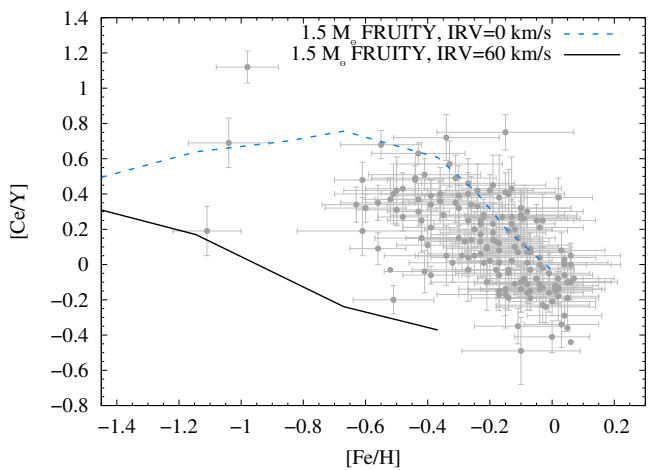}
      \caption{Comparison for [Ce/Y] between the Ba stars and the 1.5 \(M_\odot\) rotating and non-rotating (IRV = initial rotational velocity) FRUITY models that achieve [s/Fe]>0.25. The dots without error bars represent stars for which there are less than 3 lines for one of the elements.}
         \label{fig:rot}
   \end{figure}

The transport of angular momentum in rotating stars has received much attentions in the past decade in relation to asteroseismology observations from the Kepler satellite. These have allowed us to infer the rotation of the stellar core as stars evolve from the main sequence onto the giant branch and demonstrate that the cores of red giant stars rotate much slower than expected by models that do not include any coupling of the faster-rotating, contracting core with the slower-rotating, expanding
envelope \citep{mosser12,deheuvels15,gehan18}. Rotational rates of white dwarfs also show that they rotate slower than expected \citep{suijs08,hermes17}. 
On the other hand, we cannot derive the core rotation for AGB stars directly because the asteroseismology observations that would allow us to do that are expected to be at low-frequency, and their usage is hampered by the frequency resolution determined by the limited length of the available Kepler observations, and by instrumental effects \citep{mosser13}. 

Regarding the $s$ process, stellar rotation and the ensuing difference in the angular momentum between the core and the envelope when the star becomes a giant has been demonstrated to drive mixing inside the \iso{13}C pocket during the neutron flux on the AGB phase and effectively diminish the neutron exposure \citep{herwig03,siess04,piersanti13}. This is because 
the partial mixing of protons from the envelope that results in the formation of the \iso{13}C pocket also produces an adjacent \iso{14}N-rich pocket \citep{goriely00,lugaro03,cristallo09,buntain17}. Rotational mixing, if it occurs, carries \iso{14}N into the \iso{13}C pocket, and the \iso{14}N(n,p)\iso{14}C reaction \citep{wallner16} effectively captures the free neutrons. Rotation could thus represent a second parameter that varies the $s$-process distribution at any given metallicity.  
   
In Fig.~\ref{fig:rot} we compare the Ba star data to 1.5 \msun\ FRUITY models computed with and without the inclusion of an initial rotational velocity (IRV) of 60 km/s \citep{piersanti13}. This is a typical value for stars of this mass, while stars of higher mass are known to initially rotate even faster, $>$100 km/s \citep{stauffer86,nielsen13}. Due to the mixing of \iso{14}N into the \iso{13}C-rich region, in the rotating models the neutron exposure in the \iso{13}C pocket is much lower than in the non-rotating models and results in much lower [Ce/Y] ratios. These are not seen in the bulk of the Ba stars, which indicates that strong mixing within the \iso{13}C pocket of their AGB companions should not occur. 

Rotating models for the metallicity range of the Ba star sample plotted in Fig.~\ref{fig:rot} are not yet available for masses different from 1.5 \msun.
For stars of masses up to 3 \msun\ several rotating models have been published at solar metallicity: 2 \msun\ stars by \citet{piersanti13}, and 3 \msun\ stars by \citet{herwig03} and \citet{siess04}. While the different codes differ in the details of the final results, the qualitative result is the same as for the 1.5 \msun\ model plotted in Fig.~\ref{fig:rot}: rotation decreases the 
efficiency of the \iso{13}C neutron source and, in turn, the [Ce/Y] ratio. 
Furthermore, as mentioned above, stars of mass above 1.5 \msun\ are 
observed to have initial rotational velocities above 100 km/s, and, for a 
given set of model inputs, the effect of rotational mixing increases with 
the initial rotational velocity \citep{piersanti13}.
Also asteroseismology observations of the slow down of the core of giant 
stars and of the rotational velocities of WDs extend to stars of initial 
mass 2.5--3 \msun\ \citep{mosser12,hermes17,gehan18}. We derive that in 
this range of mass our conclusion will hold.

On the other hand, asteroseismologic evidence is not readily available on the evolution of the angular momentum in giant stars of mass above 3 \msun. These are relatively rare in the Kepler field of view \citep{hekker11} due to the initial mass function, their shorter red giant phase ($<$100 Myr), and the fact that the field of view was shifted out of the galactic plane. One classical Cepheid in the Kepler field \citep[V1154 Cyg,][]{derekas17} has an estimated mass of 4.5 \msun. However, we do not see solar-like oscillations in this star probably because the large-amplitude violent pulsations hamper the development of observable turbulence-driven oscillations in the thin convective layers. Among the sample of \citet{hermes17}, only three WDs have progenitor mass between 3 and 4 \msun. The most massive rotates faster than any other pulsating WDs. This may indicate a link between higher mass and faster rotation, but more data are required to confirm this trend. In this mass range above 3 \msun, models for the $s$ process including stellar rotation are also still missing.

However, based on the nucleosynthesis evidence detailed below, we can conclude that AGB stars with initial mass much higher than 3 \msun\ cannot be responsible for the bulk of the Ba stars observations. First, both observations \citep{garcia13} and models \citep{goriely04,cristallo15} show that the importance of the \iso{13}C 
neutron source decreases with the stellar mass. In fact, the FRUITY 
models of 4 and 5 \msun\ also presented here do not produce enough 
$s$-process enhancements to cover the bulk of the Ba stars 
observations (see Fig. \ref{fig:cefe}). Second, the \iso{22}Ne source can be more significantly activated in these models and help reaching the 
required $s$-process enhancements (for example, as in the 4 and 4.5 \msun\ 
Monash models presented here), however, as discussed above, this neutron 
source results in high neutron densities and the efficient production of 
Rb, with [Rb/Zr] and [Rb/Y] ratios always
above $-$0.25 dex (see Tables \ref{tab:FRUITY} and \ref{tab:monash}). This is in contrast with observations, which show that Ba stars have [Rb/Zr] and [Rb/Y]$<-$0.4: Fig.~7 of \citet{abia98}, Sec.~6.2 of \citet{busso99}, and Fig.~18 of \citet{karinkuzhi18b}. Observations 
of massive AGB stars have confirmed that these stars in fact produce Rb \citep{garcia06,garcia09}, although a quantitative mismatch with the 
models is still present \citep{vanraai12,karakas12} and currently being 
investigated \citep{perez17}.

A braking mechanism to slow down the core is already urgently looked for on the basis of the asteroseismology data \citep{eggenberger17}. Magnetic field \citep{cantiello14}, gravity waves \citep{fuller14}, and mixed modes, i.e, g-modes excited in the core coupled to p-modes present in the atmosphere \citep{belkacem15} have been considered so far as possibilities. It could be stressed that a lower injection of initial angular momentum may improve the situation. However, \cite{piersanti13} already demonstrated that the introduction of a strong coupling between core and envelope prior to the AGB phase (needed to reproduce observations) would cancel any mixing induced by rotation.
The observations of Ba stars provide an independent and complementary constraint for the presence of a mechanism for transporting angular momentum. Further, they indicate that such a mechanism should transport the angular momentum, but not the chemical species.

\section{Conclusion}

Based on the large (169 stars) data set of observations of $s$-process elements in Ba stars of deC16, we have performed a new comparison between data and model predictions including calculation of more accurate error bars for the ratios of hs elements (Ce and Nd, belonging to the second $s$-process peak) to ls elements (Y and Zr, belonging to the first $s$-process peak).

 We have compared the results to two sets of models (FRUITY and Monash) for the $s$ process in the AGB star of masses between 1.5 and 4 \msun\ and [Fe/H] between 0 and $-$1.2 believed to have transferred the $s$-process elements onto the companion Ba star. Our main results are as follows:

\begin{enumerate}

    \item In our analysis we excluded La because we found that the La abundance may be overestimated in some Ba stars. The [La/Fe] ratios reaches up to $\simeq$ 2.5 dex, well above the [Ce/Fe] and [Nd/Fe] ratios, where La, Ce, and Nd all belong to the second peak of the $s$-process elements and are necessarily produced by a similar factor. The fact that very strong La lines are present in the sample spectra makes the abundance determination of La in these stars unreliable.
    
    \item All the computed ratios [Ce/Y], [Ce/Zr], [Nd/Y], [Nd/Zr] show a clear trend of increasing with decreasing the stellar metallicity. This is in very good agreement with the models, and confirms that the main neutron source in AGB stars, the \iso{13}C nuclei in the \iso{13}C pocket, are of primary origin, i.e., their abundance is independent on the metallicity of the star. 
    
    \item At any given metallicity a spread of roughly a factor of 3 is shown by the data. This could be explained by a variety of processes (related for example to the stellar mass, overshoot, rotation, magnetic fields), however, the uncertainties are currently too large to allow us to identify which of these effects plays the main role.
    
    \item Rotating low-mass AGB models (with initial mass < 3 \msun) produce [Ce/Y], [Ce/Zr], [Nd/Y], [Nd/Zr] ratios much lower than those observed in Ba stars. This requires the existence of a mechanism for the transport of angular momentum, but not of chemical species, to be active in giant stars, in agreement with independent constraints from Kepler asteroseismology observations on the rotational velocities of the cores of giant stars and of WDs. Rotating more massive AGB models (with initial mass > 3 \msun) are not yet available, however, such stars cannot be invoked as responsible for producing the bulk of the observed Ba star data. This is because their \iso{13}C pockets are too small to result in $s$-process enhancements high enough to match the observations and/or the more significantly activated neutron source \iso{22}Ne results in production of Rb higher than observed \citep{abia98,karinkuzhi18b}.

\end{enumerate}

Future work involves the analysis of other elements in the same set of Ba stars, from C and O to other heavy elements, such as Rb (also to further confirm that the \iso{22}Ne is not the main source for the $s$ process) and Pb. Furthermore, the Ba star set of deC16 should be compared to the set of \citet{escorza17} -- there are roughly 70 stars in common -- to associate a mass to each star and verify if the composition of each Ba star can be matched by models of the appropriate mass. The Ba star observations also need to be compared to observations of other $s$-process enhanced stars, from C stars \citep{abia02}, to post-AGB stars \citep{desmedt16} as well as CH and CEMP stars \citep{lugaro12,abate15b,abate15a,cristallo16}. 
Future work also involves the detailed analysis of the effect of upcoming new experimental data on the neutron-capture cross sections of the isotopes of interest here.
Finally, meteoritic stardust SiC grains from C-rich AGB stars also show the clear indication of the $s$-process in their parent stars \citep{lugaro03b,liu14b,liu14a,liu15,lugaro18}. They should be also considered together with spectroscopic observations to produce a complete picture of the $s$-process in AGB stars.



%

\begin{acknowledgements}

We thank the referee for comments that helped to improve the discussion and conclusions of this paper.
B.Cs., M.L., L.M., E.P., and R.Sz. acknowledge the support of the Hungarian Academy of Sciences via the Lend\"ulet project LP2014-17.
B.Cs. and V.d'O. acknowledge support from the “ChETEC” COST Action (CA16117), supported by COST (European Cooperation in Science and Technology)”.
This research was supported in part by the National Science Foundation under Grant No. PHY-1430152 (JINA Center for the Evolution of the Elements).
This project was supported by K-115709 grant of the National Research, Development and Innovation Fund of Hungary, financed under the K\_16 funding scheme. This project has also been supported by the Lend\"ulet Program  of the Hungarian Academy of Sciences, project No. LP2018-7/2018.
The research leading to these results have been supported by Research, Development and Innovation Office (NKFIH) grants K-115709, PD-116175, and PD-121203. L.M. and E.P. were supported by the J\'anos Bolyai Research Scholarship of the Hungarian Academy of Sciences.
A.I.K. acknowledges financial support from the Australian Research Council (DP170100521). M.P. acknowledge the support of STFC through the University of Hull Consolidated Grant ST/R000840/1, and  ongoing resource allocations on the University of Hulls High Performance Computing Facility viper.
\end{acknowledgements}


\bibliographystyle{astron}
\bibliography{bibliography.bib}

\newpage
\onecolumn
\begin{appendix}
\section{}

\begin{longtable}{lcccccccccc}
\caption{The abundance ratios and the calculated errors for the hs-type to ls-type elements for the sample stars} \\

\hline \hline
\textbf{name} & \textbf{[Fe/H]} &	\textbf{e\_[Fe/H]}	& \textbf{[Ce/Y]} & \textbf{e\_[Ce/Y]} & \textbf{[Ce/Zr]} & \textbf{e\_[Ce/Zr]} & \textbf{[Nd/Y]} & \textbf{e\_[Nd/Y]} & \textbf{[Nd/Zr]} & \textbf{e\_[Nd/Zr]} \\
\hline
\endfirsthead
\multicolumn{11}{c}
{\tablename\ \thetable\ -- \textit{Continued from previous page}} \\
\hline \hline
\textbf{name} & \textbf{[Fe/H]} &	\textbf{e\_[Fe/H]}	& \textbf{[Ce/Y]} & \textbf{e\_[Ce/Y]} & \textbf{[Ce/Zr]} & \textbf{e\_[Ce/Zr]} & \textbf{[Nd/Y]} & \textbf{e\_[Nd/Y]} & \textbf{[Nd/Zr]} & \textbf{e\_[Nd/Zr]} \\
\hline
\endhead
\endfoot
\hline
\endlastfoot

BD-08 3194  &  -0.10  &  0.16  &  0.32  &  0.11  &  0.32  &  0.24  & 0.40 & 0.15  &  0.40  &  0.22 \\
BD-09 4337  &  -0.24  &  0.21  &  0.33  &  -  &  -0.07  &  0.25  & 0.45 & -  &  0.05  &  0.23 \\
BD-14 2678  &  +0.01  &  0.12  &  -0.15  &  0.07  &  0.02  &  0.22  & -0.15 & 0.08  &  0.02  &  0.24 \\
CD-27 2233  &  -0.25  &  0.18  &  0.05  &  0.10  &  0.21  &  0.24  & 0.07 & 0.14  &  0.23  &  0.22 \\
CD-29 8822  &  +0.04  &  0.15  &  0  &  0.07  &  0.25  &  0.22  & -0.15 & 0.07  &  0.10  &  0.24 \\
CD-30 8774  &  -0.11  &  0.14  &  -0.35  &  0.10  &  0.11  &  0.24  & -0.42 & 0.14  &  0.04  &  0.22 \\
CD-38 585  &  -0.52  &  0.09  &  0.37  &  0.12  &  0.47  &  0.24  & 0.50 & 0.16  &  0.60  &  0.22 \\
CD-42 2048  &  -0.23  &  0.16  &  0.17  &  0.15  &  0.16  &  0.28  & 0.27 & 0.16  &  0.26  &  0.23 \\
CD-53 8144  &  -0.19  &  0.15  &  0.14  &  0.08  &  0.17  &  0.25  & 0.17 & 0.11  &  0.20  &  0.22 \\
CD-61 1941  &  -0.20  &  0.14  &  0.42  &  0.09  &  0.47  &  0.24  & 0.35 & 0.13  &  0.40  &  0.22 \\
CPD-62 1013  &  -0.08  &  0.14  &  0.07  &  0.09  &  0.07  &  0.25  & 0.02 & 0.10  &  0.02  &  0.28 \\
CPD-64 3333  &  -0.10  &  0.18  &  0.28  &  0.14  &  0.29  &  0.25  & 0.28 & 0.15  &  0.29  &  0.22 \\
HD 4084  &  -0.42  &  0.15  &  0.15  &  0.08  &  -0.18  &  0.24  & 0.05 & 0.13  &  -0.28  &  0.22 \\
HD 5424  &  -0.41  &  0.18  &  0.51  &  0.09  &  0.70  &  0.25  & 0.44 & 0.12  &  0.63  &  0.22 \\
HD 5825  &  -0.48  &  0.08  &  0.27  &  0.08  &  0.47  &  0.22  & 0.08 & 0.07  &  0.28  &  0.24 \\
HD 15589  &  -0.27  &  0.15  &  0.46  &  0.14  &  0.30  &  0.25  & 0.54 & 0.16  &  0.38  &  0.22 \\
HD 20394  &  -0.22  &  0.12  &  0.28  &  0.07  &  0.14  &  0.21  & 0.30 & 0.08  &  0.16  &  0.24 \\
HD 21989  &  -0.14  &  0.17  &  -0.05  &  0.13  &  0.19  &  0.28  & -0.15 & 0.14  &  0.09  &  0.22 \\
HD 22285  &  -0.60  &  0.13  &  0.32  &  0.08  &  0.29  &  0.24  & 0.27 & 0.13  &  0.24  &  0.22 \\
HD 22772  &  -0.17  &  0.13  &  0.01  &  0.08  &  0.28  &  0.25  & -0.07 & 0.12  &  0.20  &  0.22 \\
HD 24035  &  -0.23  &  0.15  &  0.23  &  0.10  &  0.38  &  0.26  & 0.23 & 0.13  &  0.38  &  0.24 \\
HD 29370  &  -0.25  &  0.16  &  0.20  &  0.10  &  0.32  &  0.26  & 0.17 & 0.12  &  0.29  &  0.22 \\
HD 29685  &  -0.07  &  0.14  &  -0.03  &  0.07  &  0.20  &  0.24  & -0.15 & 0.12  &  0.08  &  0.22 \\
HD 30240  &  +0.02  &  0.15  &  -0.14  &  0.09  &  0.12  &  0.23  & -0.21 & 0.06  &  0.05  &  0.24 \\
HD 30554  &  -0.12  &  0.14  &  0.10  &  0.08  &  0.35  &  0.25  & -0.05 & 0.12  &  0.20  &  0.22 \\
HD 32712  &  -0.24  &  0.16  &  0.42  &  0.15  &  0.60  &  0.28  & 0.45 & 0.16  &  0.63  &  0.23 \\
HD 32901  &  -0.44  &  0.14  &  0.49  &  0.14  &  0.61  &  0.28  & 0.50 & 0.15  &  0.62  &  0.23 \\
HD 35993  &  -0.05  &  0.12  &  0.25  &  0.09  &  0.23  &  0.22  & 0.18 & 0.08  &  0.16  &  0.24 \\
HD 36650  &  -0.28  &  0.13  &  0.13  &  0.07  &  0.22  &  0.24  & 0.02 & 0.12  &  0.11  &  0.22 \\
HD 38488  &  +0.05  &  0.10  &  -0.19  &  0.19  &  0.08  &  0.29  & -0.08 & 0.18  &  0.19  &  0.23 \\
HD 40430  &  -0.23  &  0.13  &  0.13  &  0.08  &  0.31  &  0.25  & -0.02 & 0.11  &  0.16  &  0.22 \\
HD 43389  &  -0.50  &  0.17  &  0.42  &  0.19  &  0.02  &  0.28  & 0.55 & 0.20  &  0.15  &  0.23 \\
HD 49641  &  -0.30  &  0.17  &  0.15  &  0.12  &  0.51  &  0.28  & 0.25 & 0.13  &  0.61  &  0.22 \\
HD 51959  &  -0.10  &  0.15  &  0.26  &  0.10  &  0.11  &  0.22  & 0.20 & 0.10  &  0.05  &  0.24 \\
HD 58368  &  +0.04  &  0.14  &  0.01  &  0.06  &  0.26  &  0.22  & -0.16 & 0.06  &  0.09  &  0.24 \\
HD 59852  &  -0.22  &  0.10  &  0.01  &  0.08  &  0.14  &  0.21  & -0.10 & 0.08  &  0.03  &  0.24 \\
HD 61332  &  +0.07  &  0.13  &  -0.08  &  0.08  &  0.04  &  0.25  & -0.05 & 0.12  &  0.07  &  0.22 \\
HD 64425  &  +0.06  &  0.16  &  0.05  &  0.07  &  0.10  &  0.25  & 0.09 & 0.12  &  0.14  &  0.22 \\
HD 66291  &  -0.31  &  0.15  &  0.35  &  0.11  &  0.52  &  0.28  & -0.15 & 0.13  &  0.02  &  0.23 \\
HD 67036  &  -0.41  &  0.13  &  -0.04  &  0.12  &  0.15  &  0.28  & 0.02 & 0.14  &  0.21  &  0.22 \\
HD 71458  &  -0.03  &  0.10  &  -0.02  &  0.12  &  0.15  &  0.28  & -0.01 & 0.13  &  0.16  &  0.22 \\
HD 74950  &  -0.21  &  0.13  &  0.09  &  0.24  &  0.14  &  0.29  & 0.08 & 0.33  &  0.13  &  0.32 \\
HD 82221  &  -0.21  &  0.18  &  -0.10  &  0.16  &  0.01  &  0.28  & -0.13 & 0.16  &  -0.02  &  0.23 \\
HD 83548  &  +0.03  &  0.14  &  0.08  &  0.10  &  0.06  &  0.22  & -0.07 & 0.10  &  -0.09  &  0.25 \\
HD 84610  &  0.00  &  0.14  &  -0.21  &  0.08  &  0.02  &  0.25  & -0.25 & 0.11  &  -0.02  &  0.22 \\
HD 84678  &  -0.13  &  0.16  &  0.43  &  0.18  &  0.31  &  0.28  & 0.65 & 0.19  &  0.53  &  0.23 \\
HD 88035  &  -0.10  &  0.18  &  0.27  &  0.06  &  0.22  &  0.24  & 0.19 & 0.11  &  0.14  &  0.22 \\
HD 88562  &  -0.27  &  0.15  &  0.25  &  0.18  &  0.10  &  0.29  & 0.31 & 0.18  &  0.16  &  0.24 \\
HD 89175  &  -0.55  &  0.13  &  0.68  &  0.08  &  0.66  &  0.25  & 0.70 & 0.13  &  0.68  &  0.22 \\
HD 91208  &  +0.05  &  0.14  &  -0.10  &  0.10  &  0.23  &  0.23  & -0.27 & 0.07  &  0.06  &  0.24 \\
HD 91979  &  -0.11  &  0.12  &  0.15  &  0.07  &  0.24  &  0.24  & -0.05 & 0.12  &  0.04  &  0.22 \\
HD 92626  &  -0.15  &  0.22  &  0.75  &  0.10  &  0.53  &  0.25  & 0.70 & 0.14  &  0.48  &  0.22 \\
HD 105902  &  -0.18  &  0.17  &  0.13  &  0.13  &  0.17  &  0.25  & 0.15 & 0.15  &  0.19  &  0.22 \\
HD 107264  &  -0.19  &  0.17  &  0.45  &  0.17  &  0.16  &  0.28  & 0.33 & 0.18  &  0.04  &  0.23 \\
HD 107541  &  -0.63  &  0.11  &  0.34  &  0.10  &  0.52  &  0.24  & 0.30 & 0.07  &  0.48  &  0.25 \\
HD 110483  &  -0.04  &  0.14  &  0.21  &  0.07  &  0.32  &  0.24  & 0.05 & 0.12  &  0.16  &  0.22 \\
HD 110591  &  -0.56  &  0.12  &  0.35  &  0.10  &  0.61  &  0.25  & 0.18 & 0.14  &  0.44  &  0.22 \\
HD 111315  &  +0.04  &  0.09  &  -0.06  &  0.08  &  0.30  &  0.25  & -0.19 & 0.16  &  0.17  &  0.25 \\
HD 113291  &  -0.02  &  0.16  &  -0.24  &  0.09  &  0.18  &  0.24  & -0.30 & 0.14  &  0.12  &  0.22 \\
HD 116869  &  -0.36  &  0.12  &  0.36  &  0.11  &  0.59  &  0.24  & 0.22 & 0.15  &  0.45  &  0.22 \\
HD 119185  &  -0.43  &  0.10  &  0.30  &  0.08  &  0.39  &  0.25  & 0.14 & 0.11  &  0.23  &  0.22 \\
HD 120571  &  -0.39  &  0.09  &  0.21  &  0.16  &  0.43  &  0.28  & 0.20 & 0.17  &  0.42  &  0.23 \\
HD 120620  &  -0.14  &  0.18  &  0.40  &  0.14  &  0.45  &  0.24  & 0.08 & 0.11  &  0.13  &  0.24 \\
HD 122687  &  -0.07  &  0.13  &  0.09  &  0.09  &  0.40  &  0.23  & -0.05 & 0.08  &  0.26  &  0.24 \\
HD 123396  &  -1.04  &  0.13  &  0.69  &  0.14  &  0.56  &  0.28  & 0.65 & 0.14  &  0.52  &  0.23 \\
HD 123701  &  -0.44  &  0.09  &  0.48  &  0.08  &  0.26  &  0.22  & 0.42 & 0.08  &  0.20  &  0.24 \\
HD 123949  &  -0.09  &  0.18  &  0.26  &  0.19  &  0.46  &  0.28  & 0.29 & 0.19  &  0.49  &  0.22 \\
HD 126313  &  -0.10  &  0.16  &  0.32  &  0.08  &  0.43  &  0.25  & 0.12 & 0.12  &  0.23  &  0.22 \\
HD 130255  &  -1.11  &  0.11  &  0.19  &  0.14  &  0.15  &  0.28  & 0.04 & 0.15  &  0  &  0.22 \\
HD 131670  &  -0.04  &  0.15  &  0  &  0.08  &  0.06  &  0.24  & -0.19 & 0.12  &  -0.13  &  0.22 \\
HD 136636  &  -0.04  &  0.18  &  0.04  &  0.06  &  0.25  &  0.24  & 0.07 & 0.12  &  0.28  &  0.22 \\
HD 142751  &  -0.10  &  0.13  &  -0.14  &  0.13  &  0.04  &  0.28  & -0.21 & 0.14  &  -0.03  &  0.23 \\
HD 143899  &  -0.27  &  0.12  &  0.04  &  0.11  &  0.33  &  0.22  & -0.15 & 0.10  &  0.14  &  0.24 \\
HD 147884  &  -0.09  &  0.15  &  -0.08  &  0.09  &  0.13  &  0.22  & -0.23 & 0.09  &  -0.02  &  0.24 \\
HD 148177  &  -0.15  &  0.15  &  -0.14  &  0.22  &  -0.07  &  0.28  & -0.14 & 0.23  &  -0.07  &  0.23 \\
HD 154430  &  -0.36  &  0.19  &  0.40  &  0.15  &  0.36  &  0.28  & 0.38 & 0.16  &  0.34  &  0.23 \\
HD 162806  &  -0.26  &  0.17  &  0.14  &  0.12  &  0.31  &  0.28  & 0.02 & 0.12  &  0.19  &  0.22 \\
HD 168214  &  -0.08  &  0.10  &  -0.16  &  0.10  &  -0.10  &  0.22  & -0.25 & 0.09  &  -0.19  &  0.24 \\
HD 168560  &  -0.13  &  0.13  &  0.02  &  0.14  &  0.25  &  0.28  & -0.02 & 0.15  &  0.21  &  0.23 \\
HD 168791  &  -0.23  &  0.17  &  0.06  &  0.13  &  0.03  &  0.29  & 0.04 & 0.14  &  0.01  &  0.24 \\
HD 176105  &  -0.14  &  0.12  &  -0.03  &  0.12  &  0.07  &  0.28  & -0.10 & 0.13  &  0  &  0.23 \\
HD 177192  &  -0.17  &  0.20  &  -0.17  &  0.09  &  -0.31  &  0.24  & -0.26 & 0.14  &  -0.40  &  0.22 \\
HD 180996  &  +0.06  &  0.15  &  -0.10  &  0.10  &  0.02  &  0.24  & -0.24 & 0.14  &  -0.12  &  0.22 \\
HD 182300  &  +0.06  &  0.16  &  0  &  0.10  &  0.28  &  0.22  & -0.08 & 0.09  &  0.20  &  0.24 \\
HD 183915  &  -0.39  &  0.14  &  0.34  &  0.14  &  0.54  &  0.28  & 0.48 & 0.14  &  0.68  &  0.23 \\
HD 187308  &  -0.08  &  0.11  &  -0.11  &  0.11  &  0.24  &  0.25  & -0.22 & 0.14  &  0.13  &  0.22 \\
HD 193530  &  -0.17  &  0.14  &  0.12  &  0.22  &  0.01  &  0.28  & -0.05 & 0.22  &  -0.16  &  0.23 \\
HD 196445  &  -0.19  &  0.17  &  0.18  &  0.17  &  0.12  &  0.29  & 0.14 & 0.16  &  0.08  &  0.23 \\
HD 199435  &  -0.39  &  0.12  &  0.39  &  0.09  &  0.50  &  0.22  & 0.40 & 0.07  &  0.51  &  0.24 \\
HD 200995  &  -0.03  &  0.17  &  0.25  &  0.15  &  0.33  &  0.29  & -0.13 & 0.13  &  -0.05  &  0.23 \\
HD 201657  &  -0.34  &  0.17  &  0.72  &  0.13  &  0.46  &  0.25  & 0.56 & 0.15  &  0.30  &  0.22 \\
HD 201824  &  -0.33  &  0.17  &  0.57  &  0.13  &  0.61  &  0.26  & 0.44 & 0.14  &  0.48  &  0.22 \\
HD 204075  &  +0.06  &  0.17  &  -0.44  &  -  &  -0.14  &  0.22  & -0.54 & -  &  -0.24  &  0.25 \\
HD 207277  &  -0.13  &  0.14  &  0.23  &  0.13  &  0.42  &  0.28  & 0.03 & 0.14  &  0.22  &  0.22 \\
HD 210709  &  -0.10  &  0.14  &  0.28  &  0.13  &  0.45  &  0.26  & 0.18 & 0.13  &  0.35  &  0.22 \\
HD 210946  &  -0.12  &  0.13  &  -0.11  &  0.08  &  0.10  &  0.24  & -0.16 & 0.13  &  0.05  &  0.22 \\
HD 211173  &  -0.39  &  0.09  &  -0.06  &  0.05  &  0.15  &  0.24  & -0.22 & 0.11  &  -0.01  &  0.21 \\
HD 211594  &  -0.43  &  0.14  &  0.37  &  0.09  &  0.39  &  0.25  & 0.29 & 0.12  &  0.31  &  0.22 \\
HD 211954  &  -0.51  &  0.19  &  0.40  &  0.17  &  0.52  &  0.29  & 0.55 & 0.15  &  0.67  &  0.22 \\
HD 214579  &  -0.26  &  0.14  &  0.34  &  0.16  &  0.10  &  0.28  & 0.29 & 0.16  &  0.05  &  0.23 \\
HD 217143  &  -0.35  &  0.17  &  0.28  &  0.13  &  0.53  &  0.28  & 0.21 & 0.14  &  0.46  &  0.23 \\
HD 217447  &  -0.17  &  0.11  &  0.08  &  0.08  &  0.41  &  0.22  & -0.09 & 0.07  &  0.24  &  0.24 \\
HD 219116  &  -0.61  &  0.09  &  0.48  &  0.10  &  0.42  &  0.24  & 0.38 & 0.14  &  0.32  &  0.22 \\
HD 223586  &  -0.08  &  0.11  &  0.01  &  0.14  &  0.07  &  0.24  & -0.06 & 0.17  &  0  &  0.22 \\
HD 223617  &  -0.18  &  0.13  &  -0.03  &  0.08  &  -0.06  &  0.24  & -0.08 & 0.12  &  -0.11  &  0.22 \\
HD 252117  &  -0.14  &  0.19  &  0.19  &  0.15  &  0.27  &  0.29  & 0.15 & 0.15  &  0.23  &  0.22 \\
HD 273845  &  -0.15  &  0.16  &  0.41  &  0.11  &  0.74  &  0.25  & 0.28 & 0.13  &  0.61  &  0.22 \\
HD 288174  &  -0.05  &  0.15  &  -0.08  &  0.08  &  0.22  &  0.25  & -0.21 & 0.11  &  0.09  &  0.22 \\
MFU 112  &  -0.43  &  0.15  &  0.63  &  0.06  &  0.36  &  0.24  & 0.40 & 0.12  &  0.13  &  0.22 \\
BD-18 821  &  -0.27  &  0.15  &  0.40  &  0.12  &  0.77  &  0.22  & 0.21 & 0.12  &  0.58  &  0.24 \\
CD-26 7844  &  +0.02  &  0.11  &  -0.13  &  0.09  &  0  &  0.22  & -0.36 & 0.08  &  -0.23  &  0.24 \\
CD-30 9005  &  +0.05  &  0.12  &  -0.10  &  0.08  &  0.16  &  0.25  & -0.29 & 0.12  &  -0.03  &  0.22 \\
CD-34 6139  &  -0.07  &  0.13  &  0.07  &  0.07  &  -0.05  &  0.25  & -0.10 & 0.12  &  -0.22  &  0.22 \\
CD-34 7430  &  +0.01  &  0.14  &  -0.12  &  0.06  &  0.05  &  0.25  & -0.27 & 0.12  &  -0.10  &  0.23 \\
CD-46 3977  &  -0.10  &  0.15  &  -0.10  &  0.03  &  0.03  &  0.24  & -0.13 & 0.12  &  0  &  0.22 \\
HD 18182  &  -0.17  &  0.10  &  -0.15  &  0.09  &  0  &  0.24  & -0.34 & 0.13  &  -0.19  &  0.22 \\
HD 18361  &  +0.01  &  0.15  &  -0.14  &  0.04  &  -0.14  &  0.24  & -0.18 & 0.12  &  -0.18  &  0.22 \\
HD 21682  &  -0.48  &  0.12  &  0.43  &  0.09  &  0.39  &  0.23  & 0.30 & 0.08  &  0.26  &  0.25 \\
HD 26886  &  -0.30  &  0.10  &  0.32  &  0.13  &  0.26  &  0.24  & 0.17 & 0.08  &  0.11  &  0.24 \\
HD 31812  &  -0.07  &  0.11  &  -0.15  &  0.07  &  -0.07  &  0.22  & -0.29 & 0.07  &  -0.21  &  0.24 \\
HD 33709  &  -0.20  &  0.14  &  0.02  &  0.11  &  0.29  &  0.23  & -0.18 & 0.08  &  0.09  &  0.24 \\
HD 39778  &  -0.12  &  0.12  &  0.20  &  0.07  &  0.33  &  0.22  & -0.04 & 0.06  &  0.09  &  0.24 \\
HD 41701  &  +0.02  &  0.13  &  -0.08  &  0.08  &  0.18  &  0.22  & -0.24 & 0.08  &  0.02  &  0.24 \\
HD 45483  &  -0.14  &  0.12  &  -0.19  &  0.07  &  0.01  &  0.24  & -0.31 & 0.12  &  -0.11  &  0.22 \\
HD 48814  &  -0.07  &  0.11  &  -0.19  &  0.07  &  0.07  &  0.24  & -0.31 & 0.12  &  -0.05  &  0.22 \\
HD 49017  &  +0.02  &  0.11  &  0.38  &  0.11  &  0.42  &  0.23  & 0.14 & 0.08  &  0.18  &  0.24 \\
HD 49661  &  -0.13  &  0.10  &  0.27  &  0.10  &  0.20  &  0.23  & 0.06 & 0.09  &  -0.01  &  0.25 \\
HD 49778  &  -0.22  &  0.12  &  0.27  &  0.10  &  0.18  &  0.22  & 0.20 & 0.11  &  0.11  &  0.24 \\
HD 50075  &  -0.16  &  0.11  &  0.27  &  0.06  &  0.45  &  0.24  & 0.11 & 0.11  &  0.29  &  0.21 \\
HD 50843  &  -0.31  &  0.13  &  0.49  &  0.13  &  0.60  &  0.25  & 0.26 & 0.15  &  0.37  &  0.22 \\
HD 53199  &  -0.23  &  0.13  &  0.25  &  0.10  &  0.23  &  0.23  & 0.09 & 0.06  &  0.07  &  0.24 \\
HD 58121  &  -0.01  &  0.13  &  -0.05  &  0.11  &  0.10  &  0.28  & -0.25 & 0.13  &  -0.10  &  0.23 \\
HD 62017  &  -0.32  &  0.14  &  0.01  &  0.11  &  0.18  &  0.22  & -0.21 & 0.10  &  -0.04  &  0.24 \\
HD 88495  &  -0.11  &  0.10  &  0.07  &  0.08  &  0.09  &  0.24  & -0.25 & 0.12  &  -0.23  &  0.22 \\
HD 90167  &  -0.04  &  0.11  &  -0.14  &  0.10  &  -0.09  &  0.23  & -0.31 & 0.10  &  -0.26  &  0.25 \\
HD 95193  &  +0.04  &  0.12  &  -0.29  &  0.09  &  -0.02  &  0.22  & -0.36 & 0.08  &  -0.09  &  0.24 \\
HD 107270  &  +0.05  &  0.17  &  -0.36  &  -  &  -0.45  &  0.22  & -0.39 & -  &  -0.48  &  0.24 \\
HD 109061  &  -0.56  &  0.09  &  0.09  &  0.09  &  0.33  &  0.24  & -0.01 & 0.14  &  0.23  &  0.22 \\
HD 113195  &  -0.15  &  0.12  &  -0.14  &  0.11  &  0.04  &  0.25  & -0.33 & 0.13  &  -0.15  &  0.22 \\
HD 115277  &  -0.03  &  0.15  &  -0.23  &  0.10  &  -0.09  &  0.25  & -0.40 & 0.13  &  -0.26  &  0.22 \\
HD 119650  &  -0.10  &  0.13  &  -0.14  &  0.13  &  0.10  &  0.28  & -0.16 & 0.15  &  0.08  &  0.23 \\
HD 134698  &  -0.52  &  0.12  &  -0.03  &  -  &  -0.06  &  0.28  & -0.16 & -  &  -0.19  &  0.23 \\
HD 139266  &  -0.27  &  0.18  &  -0.03  &  0.18  &  0.31  &  0.28  & -0.01 & 0.19  &  0.33  &  0.23 \\
HD 139409  &  -0.51  &  0.13  &  -0.20  &  0.08  &  0.01  &  0.25  & -0.40 & 0.12  &  -0.19  &  0.22 \\
HD 148892  &  -0.15  &  0.15  &  -0.17  &  0.12  &  0.12  &  0.23  & -0.18 & 0.11  &  0.11  &  0.24 \\
HD 169106  &  +0.01  &  0.12  &  -0.13  &  0.07  &  0.02  &  0.25  & -0.21 & 0.11  &  -0.06  &  0.22 \\
HD 184001  &  -0.21  &  0.14  &  0.10  &  0.13  &  0.03  &  0.22  & -0.10 & 0.13  &  -0.17  &  0.24 \\
HD 204886  &  +0.04  &  0.15  &  0.03  &  -  &  0.07  &  -  & 0.21 & -  &  0.25  &  - \\
HD 213084  &  -0.09  &  0.15  &  0.11  &  0.06  &  0.19  &  0.22  & 0.09 & 0.07  &  0.17  &  0.24 \\
HD 223938  &  -0.42  &  0.11  &  0.25  &  0.11  &  0.44  &  0.23  & 0.12 & 0.10  &  0.31  &  0.24 \\
MFU 214  &  0.00  &  0.12  &  -0.41  &  0.09  &  -0.12  &  0.24  & -0.44 & 0.13  &  -0.15  &  0.22 \\
MFU 229  &  -0.01  &  0.11  &  -0.16  &  0.11  &  -0.05  &  0.26  & -0.27 & 0.11  &  -0.16  &  0.22 \\
HD 12392  &  -0.08  &  0.18  &  0.30  &  0.12  &  0.62  &  0.25  & 0.29 & 0.15  &  0.61  &  0.22 \\
HD 17067  &  -0.61  &  0.21  &  0.19  &  0.14  &  0.19  &  0.28  & 0.25 & 0.15  &  0.25  &  0.23 \\
HD 90127  &  -0.40  &  0.10  &  0.11  &  0.14  &  0.16  &  0.28  & 0.02 & 0.15  &  0.07  &  0.23 \\
HD 102762  &  -0.17  &  0.20  &  0.38  &  0.24  &  0.37  &  0.29  & 0.45 & 0.24  &  0.44  &  0.23 \\
HD 114678  &  -0.50  &  0.13  &  0.31  &  0.09  &  0.36  &  0.22  & 0.19 & 0.08  &  0.24  &  0.24 \\
HD 180622  &  +0.03  &  0.12  &  -0.34  &  0.13  &  -0.14  &  0.28  & -0.20 & 0.15  &  0  &  0.23 \\
HD 200063  &  -0.34  &  0.20  &  0.05  &  0.17  &  0.31  &  0.30  & 0 & 0.16  &  0.26  &  0.24 \\
HD 210030  &  -0.03  &  0.11  &  -0.13  &  0.08  &  0  &  0.24  & -0.21 & 0.13  &  -0.08  &  0.22 \\
HD 214889  &  -0.17  &  0.12  &  -0.06  &  0.09  &  0.10  &  0.25  & -0.25 & 0.12  &  -0.09  &  0.22 \\
HD 215555  &  -0.08  &  0.12  &  -0.11  &  0.09  &  0.08  &  0.22  & -0.34 & 0.08  &  -0.15  &  0.24 \\
HD 216809  &  -0.04  &  0.14  &  -0.11  &  0.23  &  -0.35  &  0.28  & 0.04 & 0.24  &  -0.20  &  0.23 \\
HD 221879  &  -0.10  &  0.19  &  -0.49  &  0.19  &  -0.39  &  0.28  & -0.31 & 0.21  &  -0.21  &  0.23 \\
HD 749  &  -0.29  &  0.15  &  0.05  &  0.09  &  0.18  &  0.24  & -0.06 & 0.12  &  0.07  &  0.22 \\
HD 88927  &  +0.02  &  0.13  &  -0.18  &  0.14  &  -0.01  &  0.29  & -0.37 & 0.13  &  -0.20  &  0.22 \\
BD+09 2384  &  -0.98  &  0.10  &  1.12  &  0.09  &  0.65  &  0.26  & 0.40 & 0.11  &  -0.07  &  0.22 \\
HD 89638  &  -0.19  &  0.11  &  0.23  &  0.07  &  0.41  &  0.24  & -0.01 & 0.11  &  0.17  &  0.22 \\
HD 187762  &  -0.30  &  0.11  &  0.39  &  0.17  &  0.37  &  0.25  & 0.19 & 0.18  &  0.17  &  0.21

\end{longtable}

\end{appendix}

\end{document}